%% file: main.tex
\documentclass{aa}
\usepackage[utf8]{inputenc}
\usepackage{amsmath}
\usepackage{amssymb}
\usepackage{graphicx} 
\usepackage{epstopdf}
\usepackage{arydshln}
\usepackage{txfonts}
\input{bibdefinitions.tex}

\title{Discovery of O stars in the tidal Magellanic Bridge}
\subtitle{Stellar parameters, abundances, and feedback of the nearest metal-poor massive stars and their implication for the Magellanic System ecology}

\author{Varsha Ramachandran \and L. M. Oskinova \and W.-R. Hamann
          }
 \institute{Institut f\"ur Physik und Astronomie,
              Universit\"at Potsdam,
              Karl-Liebknecht-Str. 24/25, D-14476 Potsdam, Germany \\
              \email{varsha@astro.physik.uni-potsdam.de}
              }
\date{Received <date> / Accepted <date>}

\abstract{
The Magellanic Bridge stretching between the Small and the Large Magellanic Cloud  (SMC and LMC),  is the nearest tidally stripped intergalactic environment. The Bridge has a significantly low average metallicity of $Z\lesssim0.1Z_{\odot}$. Here we report the first discovery of O-type stars in the Magellanic Bridge.  Three massive O stars were identified thanks to the archival spectra obtained by the ESO's Very Large Telescope FLAMES instrument. We analyze the spectra of each star using the advanced non-LTE stellar atmosphere model PoWR, which provides the physical parameters, ionizing photon fluxes, and surface abundances. The ages of the newly discovered O stars suggest that star formation in the Bridge is ongoing. Furthermore, the discovery of O stars in the Bridge implies that tidally stripped galactic tails containing low density but highly dynamical gas are capable of producing massive O stars. The multi-epoch spectra indicate that all three O stars are binaries. Despite their spatial proximity to each other, these O stars are chemically distinct. One of them is a fast-rotating giant with nearly LMC-like abundances. The other two are main-sequence stars that rotate extremely slowly and are strongly metal depleted. We discover the most nitrogen-poor O star known up to date. Taking into account the previous analyses of B stars in the Bridge, we interpret the various metal abundances as the signature of a chemically inhomogeneous interstellar medium, suggesting that the Bridge gas might have accreted during multiple episodes of tidal interaction between the Clouds. Attributing the lowest derived metal content to the primordial gas, the time of initial formation of the Bridge may date back to several Gyr. Using the Gaia and Galex color-magnitude diagrams we roughly estimate the total number of O stars in the Bridge and their total ionizing radiation. Comparing with the energetics of the diffuse ISM, we find that the contribution of the hot stars to the ionizing radiation field in the Bridge is less than 10\%, and conclude that the main sources of ionizing photons are leaks from the LMC and SMC. We estimate a lower limit for the fraction of ionizing radiation that escapes from these two dwarf galaxies.

}

\begin{document}


   \maketitle

\section{Introduction}

Metallicity is an important factor deciding on the properties, evolution, and feedback from massive stars. We lack a comprehensive understanding of metal-poor massive stars and their impact on chemistry, dynamics, and evolution of their host galaxies and the ISM. Metal-poor stars are expected to be the progenitors of energetic transients such as super-luminous supernovae, long gamma-ray bursts, and coalescing massive black holes. They dominantly contributed to the cosmic re-ionization \citep{Barkana2006Sci}.
Our current observational understanding of low metallicity (Z) massive stars is primarily based on spectroscopic studies conducted in the Small Magellanic Cloud (SMC) with metal content $Z\sim0.2\,Z_{\odot}$ \citep[e.g.,][]{Bouret2013,Ramachandran2019,Dufton2019}.  
Only a few massive stars or star-forming regions at $Z\lesssim0.1Z_{\odot}$ were observed in distant ($\gtrsim$\,1 Mpc) dwarf galaxies such as Sextans\,A, SagDIG, and Leo\,P.  However, the Magellanic Bridge, stretching between the Small and the Large Magellanic Cloud (LMC), offers a unique nearby  \citep[55\,kpc,][]{Bruns2005_HI} laboratory of low metallicity.

Chemical abundances inferred from B stars as well as gas-phase suggest a mean metallicity of $Z\lesssim0.1Z_{\odot}$ and an iron abundance of  {[}Fe/H{]}$\lesssim$-1.1 \citep{rolleston_1999_chemical,dufton_2008_iron,Lehner2008}. In addition to its low Z,  the Bridge is a  tidally stripped environment of low density and highly dynamical gas,  offering a unique chance to probe star formation and feedback under such conditions.  Furthermore, studying stellar abundance patterns in this environment allows us to place constraints on metal mixing in the ISM and hence provides clues on the interaction history of the Clouds and the formation of the Bridge. 

A population of young stars was discovered throughout the Bridge \citep{Irwin1990,Bica1995} implying star formation is still occurring in this region despite the low gas density. The recent interaction between the Clouds ($<200$\,Myr) might have triggered the formation of young stars and \hi shells in the Bridge \citep{skowron2014}.
However, this young population mainly consists of B-type stars or later.  No O stars were known in the Bridge so far\footnote{An O star identified in \cite{Irwin1990} was reclassified as B-type in  \cite{Hambly1994}.}. Based on the energetics of DEM\,171, \citet{meaburn_1986_young} and \citet{Graham2001} suspect an early O-type star to be the energy source of this shell. However, there was no direct evidence. Using the {\it Spitzer} survey  in the Bridge \citet{Chen2014} found no  young stellar objects (YSO) with $M>10\,M_{\odot}$.  They could not assess if higher-mass YSOs have simply not yet formed, or such YSOs are less likely to form in the Bridge's environment. 

In this study,  we unveil the presence of three O stars along with several B stars in the western part of the Bridge using archival spectra from ESO. The spectroscopic analysis of such low metallicity massive stars is the key method to establish their parameters, evolution, and feedback in detail. Only a few O-type stars at  $Z\lesssim0.1Z_{\odot}$ were analyzed before \citep{Garcia2019}.
 
Our sample stars are associated with the \hii  regions DEM\,168 and DEM\,169 showing  \halpha  features \citep{meaburn_1986_young}, one major \hi  shell {[}MSZ2003{]}\,105 of radius 9\arcmin\,, and several small shells \citep{Muller2003}.  The presence of young O stars in the Bridge implies that star formation is still occurring in this region despite the low gas density. 

Previous spectroscopic studies were conducted to understand stellar parameters and abundances of a handful of B stars in the Bridge \citep{rolleston_1999_chemical,lee_2005_chemical,Hambly1994}.  They found an under-abundance in the light metals (C, N, O, Mg, Si) and a mean metal abundance of $\sim -1.1$\,dex lower compared to Galactic analogs. Using the UV spectra of two B stars, \citet{Lehner2001,Lehner2008}  studied the diffuse gas in the Bridge, and suggested a low gas-phase metallicity and low density. Analysis by \citet{dufton_2008_iron} also suggested a consistently lower iron abundance in B stars compared to the present-day LMC and SMC metallicity. 

The Magellanic Bridge shows the presence of hot and warm ionized gas \citep{Lehner2002,Lehner2001,D'Onghia2016ARA&A}.  \citet{Barger2013} reported that ionizing radiation from the extra-galactic background and the Milky Way is insufficient to explain the observed \halpha  flux. Other possible sources for the ionization are the hot stars within and near the Bridge, hot gas, and photons leaking from the Magellanic Clouds.  The flux of Lyman continuum photons from stars is highly dependent on their spectral type \citep{Panagia1973} and drops dramatically after B0. E.g., a B2 star produces 600 times fewer ionizing photons per second compared to an O9.  Even one O8-9 star alone can dominate the feedback in an \hii  region in the Bridge \citep{meaburn_1986_young}. Therefore, the newly discovered O stars in the Bridge should be included in studies of ionization and feedback in this region.
 
In this paper, we present a detailed spectroscopic analysis of three O stars in the Bridge. Section\,\ref{sect:spec} discusses the spectroscopic data and the multiplicity of the systems from radial velocities. Spectroscopic analyses using Potsdam Wolf–Rayet (PoWR) atmosphere models are given in Sect.\,\ref{sec:analysis}. Stellar parameters, chemical abundances, and ionizing feedback of individual stars are presented in Sect.\,\ref{sec:results}. In Sect.\,\ref{sec:impli} we discuss the impact of the O stars on the Bridge environment. The final Sect.\,\ref{sec:summary} summarizes this study. 

\section{The first discovery of O stars in the Bridge}
\label{sect:spec}
\subsection{Data}

The present study is largely based on archival spectroscopic data obtained with the multi-object spectrograph FLAMES at ESO's Very Large Telescope (078.D-0791). Among the sample of $\sim 100$ young stars in the Bridge, we identified three O stars based on the \hei/\heii line ratios. These stars add to the few sub-SMC metallicity massive stars known and are the earliest-type stars discovered in the Bridge. The present paper focuses on these three O stars, while a subsequent paper (Paper\,II) will cover the detailed analyses of the whole young stellar population (mainly B stars), along with a detailed investigation of star formation in the Bridge.

A total of eight spectra are available for each star (including multi-exposures) taken at different epochs. The spectra were taken in both high resolution (HR, $R\sim23000$) and low resolution (LR, $R\sim7500$) mode using the GIRAFFE spectrograph. Details of the observation are given in Table\,\ref{table:observation}.

\begin{table}
\caption{Details of observation} 
\label{table:observation}
\centering
\begin{tabular}{ccccc}
\hline
\hline
\noalign{\vspace{1mm}}
Date & spectral & $\lambda$&$R$ & exposure\\
& setting & [\AA] & & time [s]\\
\noalign{\vspace{1mm}}
\hline
\noalign{\vspace{1mm}}
2006-09-16 & HR04  & 4183-4395 &23000  & 2$\times$ 2775\\
2006-09-17 &HR05A & 4340-4589 & 20000 & 2$\times$ 2775\\
2006-09-29&HR06  & 4537-4761 &23000  & 1$\times$ 2775\\
2006-10-29&HR06  & 4537-4761 &23000 &2$\times$ 2775 \\
2006-11-26&LR03  & 4499-5078  &7500  & 1$\times$ 2775\\
\noalign{\vspace{1mm}}
\hline
\end{tabular}
\end{table}

In addition to the spectra, we used various photometric data from the VizieR archive to construct the spectral energy distribution (SED). NUV and FUV magnitudes are taken from the Galex-DR6 catalog \citep{Bianchi2017}. Optical ($B, V, $, $ R$ and $ G$) and infrared photometry are taken from various catalogs  \citep{Demers1991,Henden2015,GaiaDR22018,Zacharias2012,Kato2007,Gordon2011,Cutri2013}.

\begin{table}
\caption{Coordinates and proper motions of the O stars from Gaia DR2 and their spectral subtypes} 
\label{table:stars}
\centering
\setlength{\tabcolsep}{4pt}
\begin{tabular}{cccccc}
\hline\hline
\noalign{\vspace{1mm}}
star & RA & DEC & pmra & pmdec & spectral\\
 & [\degr] & [\degr] & [mas/yr] & [mas/yr] & type  \\
 \noalign{\vspace{1mm}}
 \hline
 \noalign{\vspace{1mm}}
MBO1 & 33.54942 & -74.07853 & 1.559 & -1.136 & O8 III \\
MBO2 & 32.97933 & -74.21028 & 1.443 & -1.122 & O9 V \\
MBO3 & 33.02008 & -74.19933 & 1.568 & -1.044 & O9.5 V\\
\noalign{\vspace{1mm}}
\hline
\end{tabular}
\end{table}

\subsection{The sample}

The coordinates of the sample stars and their proper motions based on Gaia DR2 are given in Table\,\ref{table:stars}. The proper motions of the stars are in good agreement with that of the Bridge population \citep{Schmidt2020,Zivick2019}. The Gaia DR2 parallaxes of the stars are negative but consistent with zero, confirming that these are distant objects.

We sub-classified the O stars based on \hei/\heii line ratios such as \ion{He}{i\,$\lambda4388$}/\,\ion{He}{ii\,$\lambda4542$},  \ion{He}{i\,$\lambda4144$}/\,\ion{He}{ii\,$\lambda4200$} and \ion{He}{i\,$\lambda4713$}/\,\ion{He}{ii\,$\lambda4686$} as well as the \ion{Si}{iii\,$\lambda4553$}/\,\ion{He}{ii\,$\lambda4542$} ratio. We labeled them by a prefix MBO (Magellanic Bridge O stars) along with a number corresponding to the ascending order of their subtype (1--3).
We classified MBO1 (alias: [DI91]\,842) as O8\,III. The luminosity-class criteria are based on the strength of \ion{He}{ii} at 4686\,\AA. Both MBO2 (alias: [DI91]\,719) and MBO3 have very similar spectra, except the \ion{Si}{iii\,$\lambda4553$}/\,\ion{He}{ii\,$\lambda4542$} ratio being slightly higher in MBO3. So, we classified MBO2 as O9\,V and MBO3 as O9-9.5\,V.

\begin{figure}
    \centering
    \includegraphics[width=0.45\textwidth]{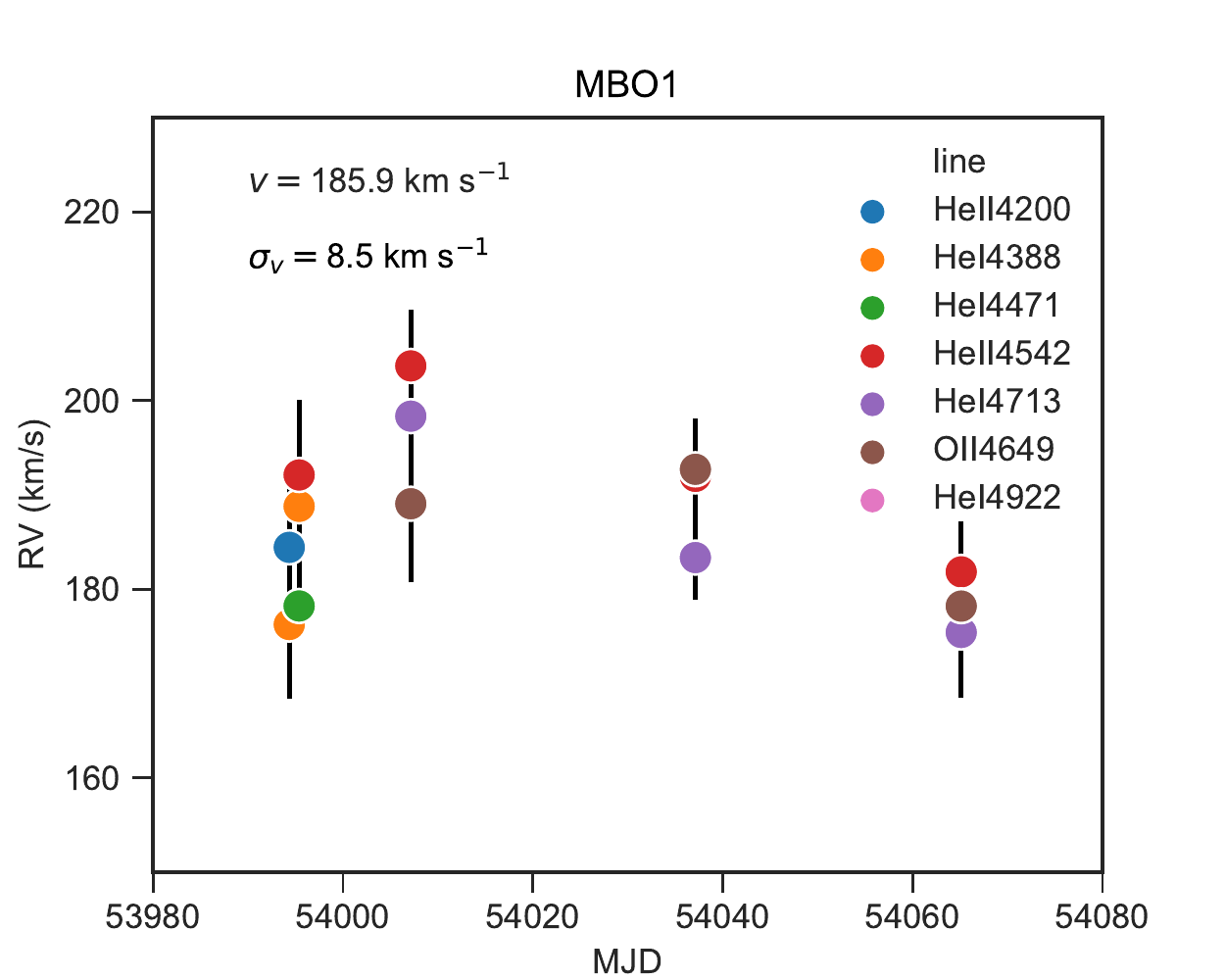}
    \includegraphics[width=0.45\textwidth]{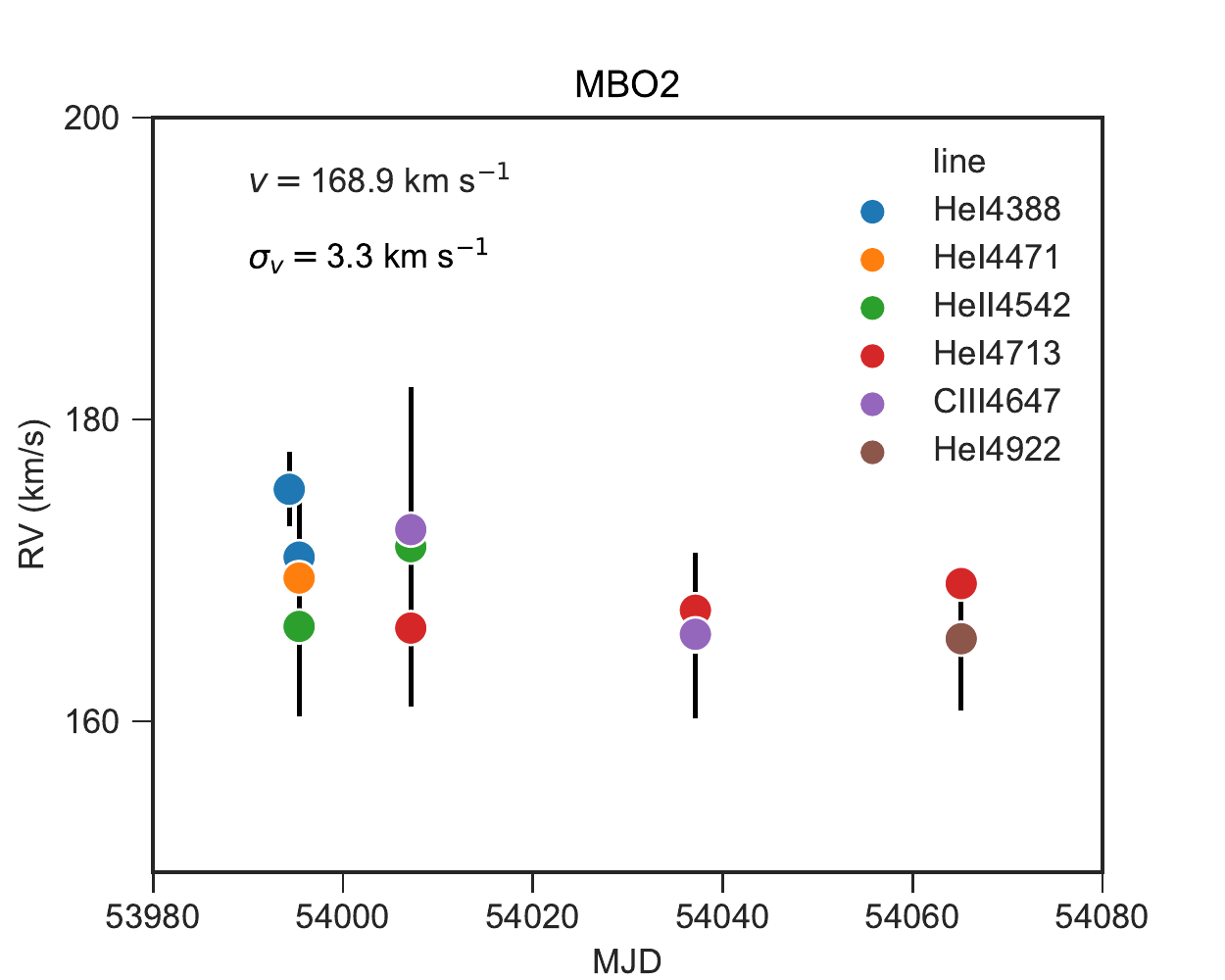}
    \includegraphics[width=0.45\textwidth]{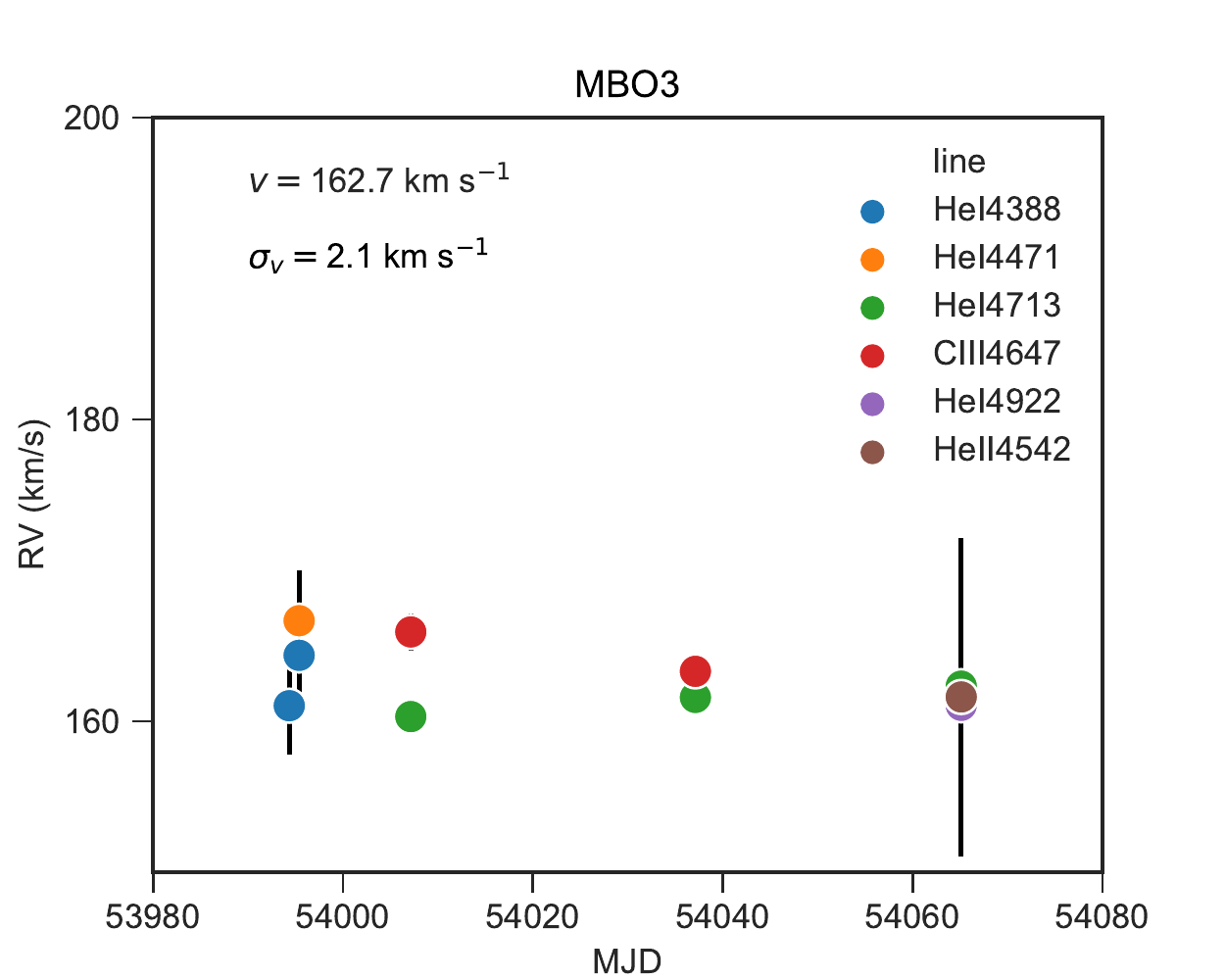}
    \caption{Radial velocity variation over time for sample O stars. Mean RV with $1\sigma$ standard deviation of the different lines are shown in different colors. Average RV over different epochs and RV dispersion of individual stars are displayed in the upper-left corner. }
    \label{fig:rv}
\end{figure} 
\subsection{Radial velocities and dispersion} 
\label{radvel}
Since all three stars have multi-epoch spectra, we checked for radial velocity (RV) variations to search for companions. Figure\,\ref{fig:rv} shows RVs of each star plotted over time in modified Julian dates (JD-2,400,000). Radial velocities were measured by fitting Gaussian profiles to various absorption lines of \hei\!, \heii\!, and  metals. 
Spectral lines used to measure RV are different in each epoch due to the different spectral settings used for observation.
All three stars were found to have only small radial velocity dispersion over a 70-day timescale ($\sigma <10\,\mathrm{km\,s^{-1}}$).

Star MBO1 shows low-amplitude RV variations but with a systematic trend. The obtained measurements yield an rms velocity dispersion of $8.5\,\mathrm{km\,s^{-1}}$ and a
peak-to-peak amplitude of about $25\,\mathrm{km\,s^{-1}}$. 
The mean uncertainty of measurements are $6.4\,\mathrm{km\,s^{-1}}$. However, while inspecting the line profiles, we could not find any peculiar lines which doesn't correspond to its spectral type. So, we suspect the star to be a  single-lined (SB1) binary.   
  
\begin{figure}
    \centering 
    \includegraphics[width=0.45\textwidth]{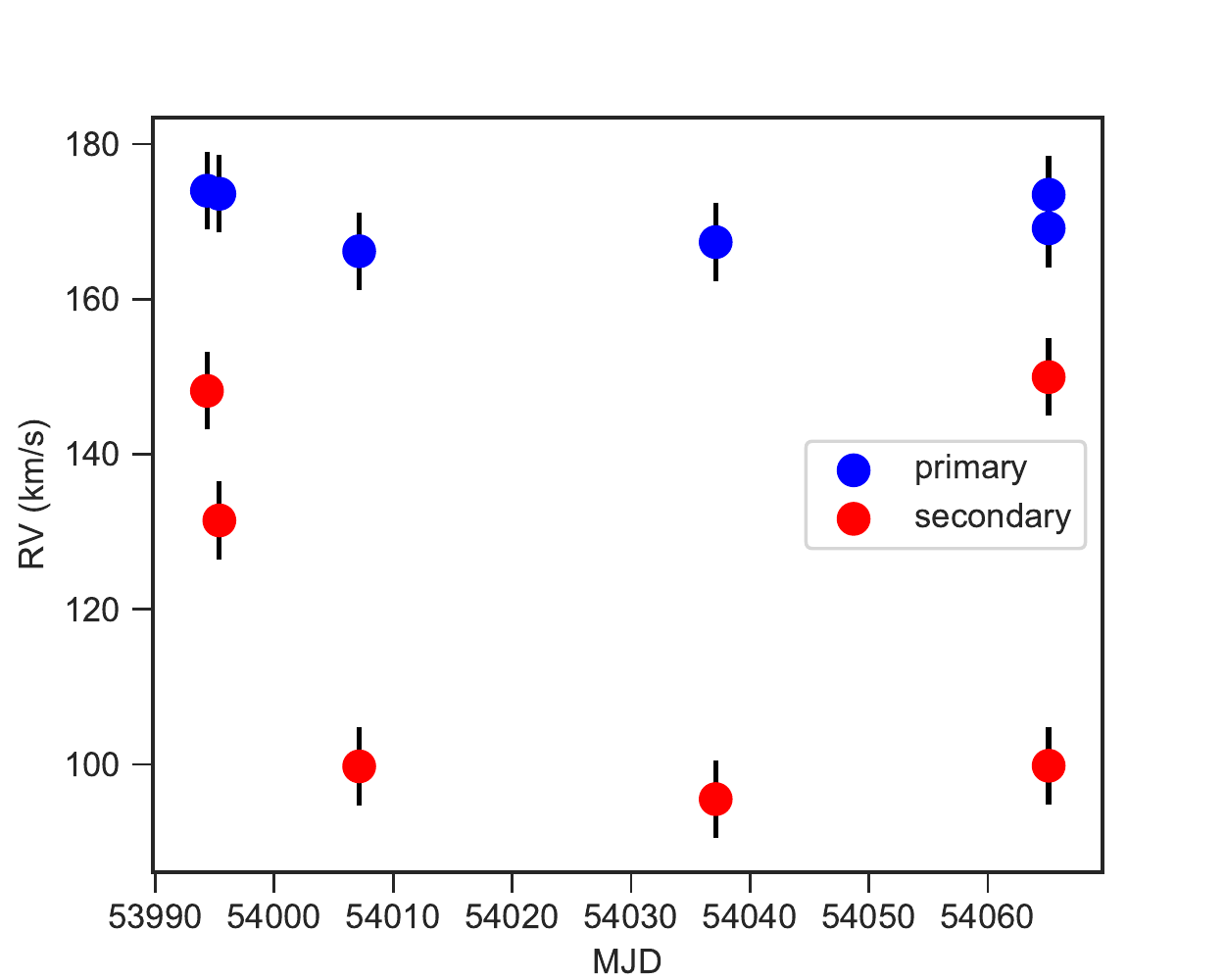}
    \includegraphics[width=0.45\textwidth]{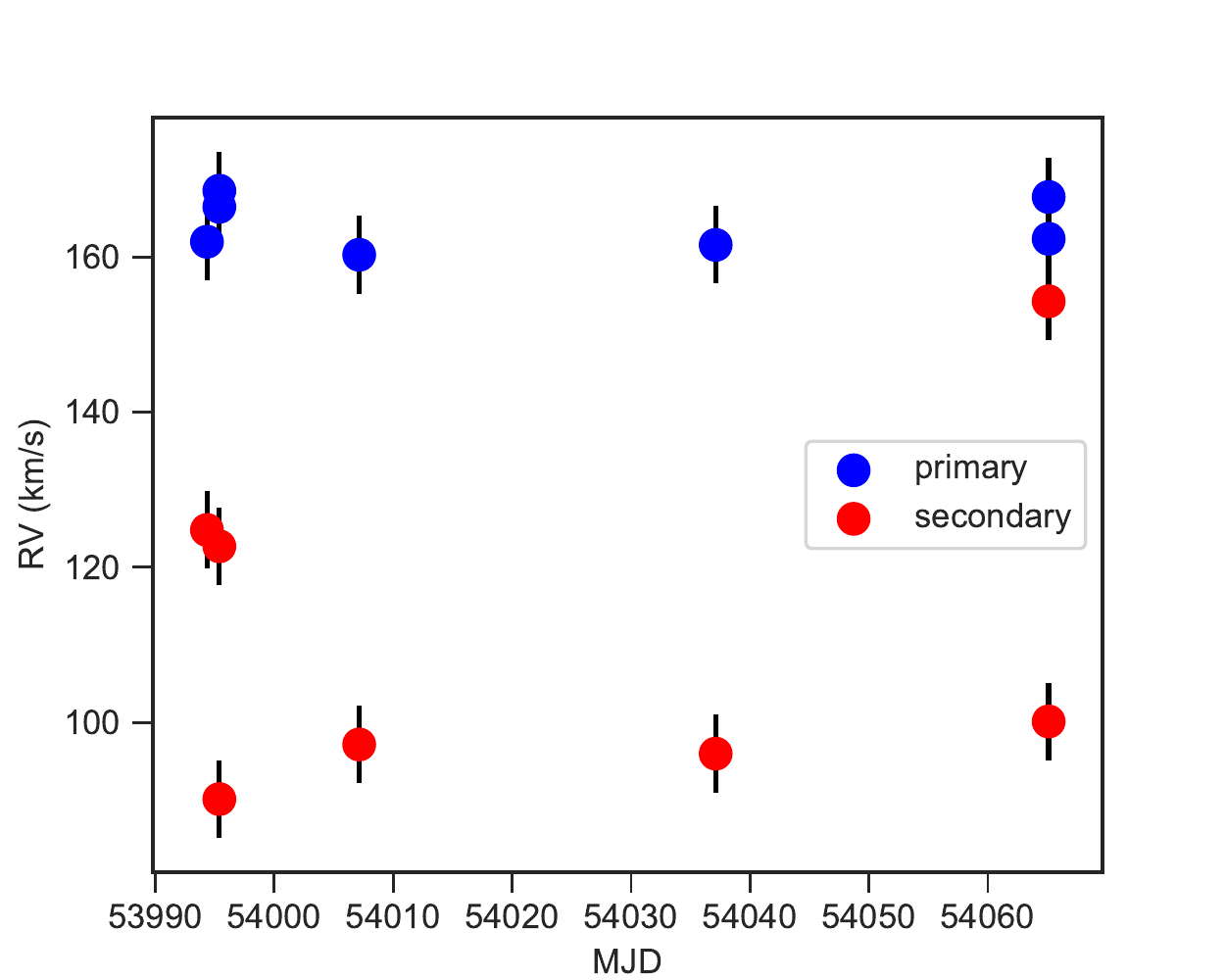}
    \caption{Radial velocity variation of primary and secondary component over time for MBO2 (upper panel) and MBO3 (lower panel). The radial velocities are measured by fitting double Gaussian profiles on \hei line profiles.}
    \label{fig:rv2}
\end{figure} 

MBO2 and MBO3 show little or no variation in RV over different epochs. In this case, rms velocity dispersion is found to be comparable to the average uncertainty of $\sim 3\,\mathrm{km\,s^{-1}}$. Despite small RV dispersion, signs of the two components are seen in the \hei lines. We fitted two Gaussian profiles per \hei line profiles. In both MBO2 and MBO3, secondary components show RVs in the range $\sim 90-150\,\mathrm{km\,s^{-1}}$ (see Fig.\,\ref{fig:rv2}).
Hence we suspect both stars to be double-lined (SB2) spectroscopic binaries.
 
The radial velocities of these O stars are in line with that of \hi gas in the Bridge \citep{Muller2003}. Their radial velocities, proper motions, and parallaxes suggest that these stars are not runaway stars from the Magellanic Clouds nor hypervelocity stars in the halo of the Milky Way.

\begin{figure*}
    \centering
    \includegraphics[width=\textwidth,trim={0 6cm 0 0}]{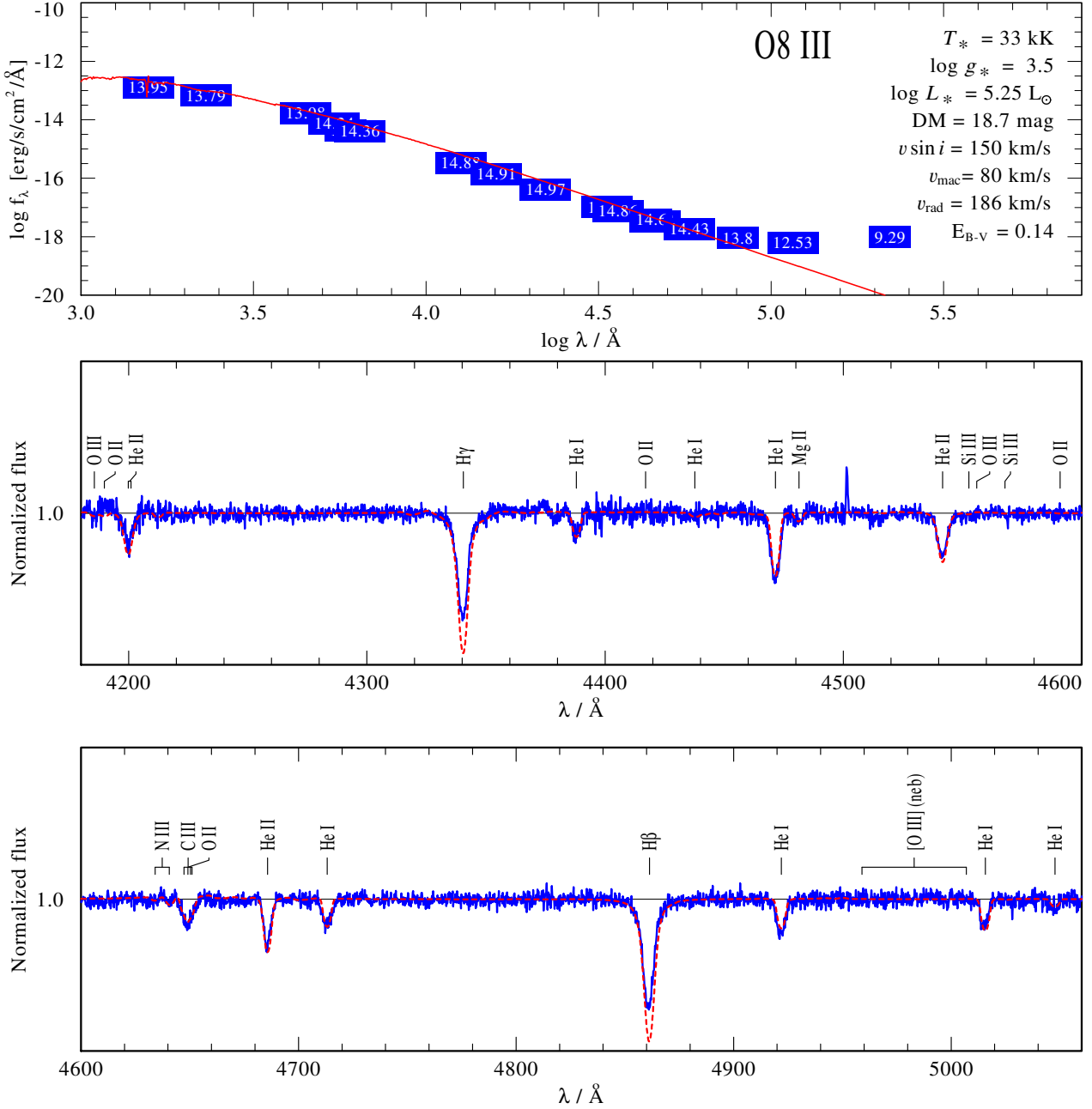}
    \caption{Spectral fit for MBO1. Upper panel: spectral energy distribution with photometry from UV, optical, and infrared bands (blue boxes) fitted to the model SED (red solid line). Lower panels: normalized VLT-FLAMES spectra (blue solid line), overplotted with the PoWR model (red dashed line). The atmospheric parameters and abundances of this best-fit model are given in Table\,\ref{table:stellarparameters} and \ref{table:abundance}.}
    \label{fig:MBO1}
\end{figure*} 

\begin{figure}
    \centering 
    \includegraphics[width=8cm,trim={1cm 19.5cm 11cm 1cm}]{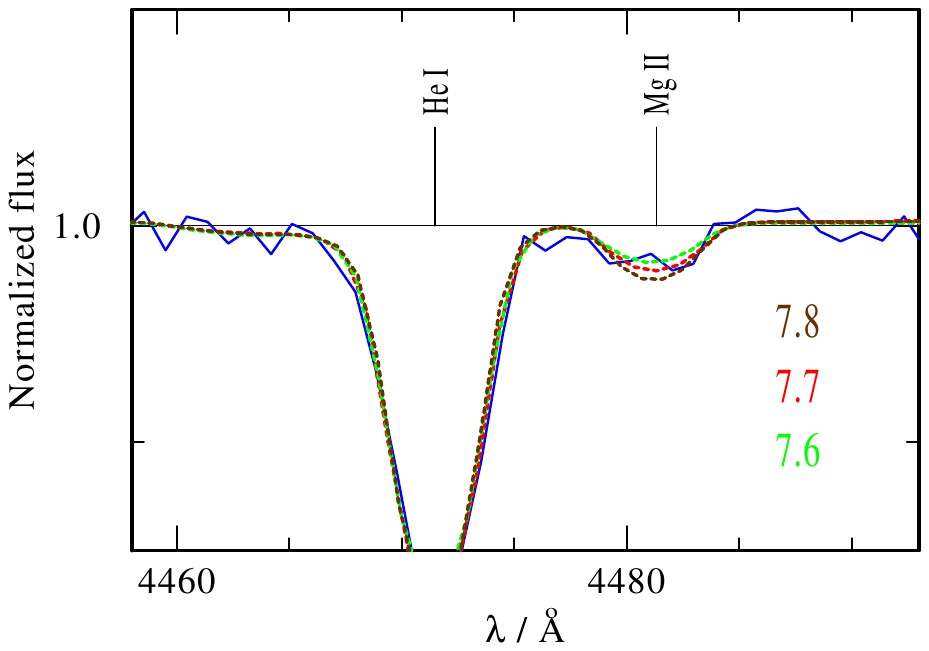}
    \caption{Comparison of \ion{Mg}{ii} line profile in MBO1 (solid blue) using synthetic models with different Mg abundances in units of $12 + \log$\,(X$_{i}/$H).}
    \label{fig:mg}
\end{figure}

\section{Analysis}
\label{sec:analysis}
To determine the parameters of the O stars, we analyzed their spectra using the Potsdam Wolf-Rayet (PoWR) model atmosphere code. PoWR is a state-of-the-art non-LTE code that can be employed for a wide range of hot stars at arbitrary metallicities \citep[e.g.][]{Hainich2014,Hainich2015,Oskinova2011,Shenar2015}. More details of the PoWR models can be found in \citet{Graefener2002,Sander2015}. 

\begin{figure*}
    \centering
    \includegraphics[width=\textwidth,trim={0 6cm 0 0}]{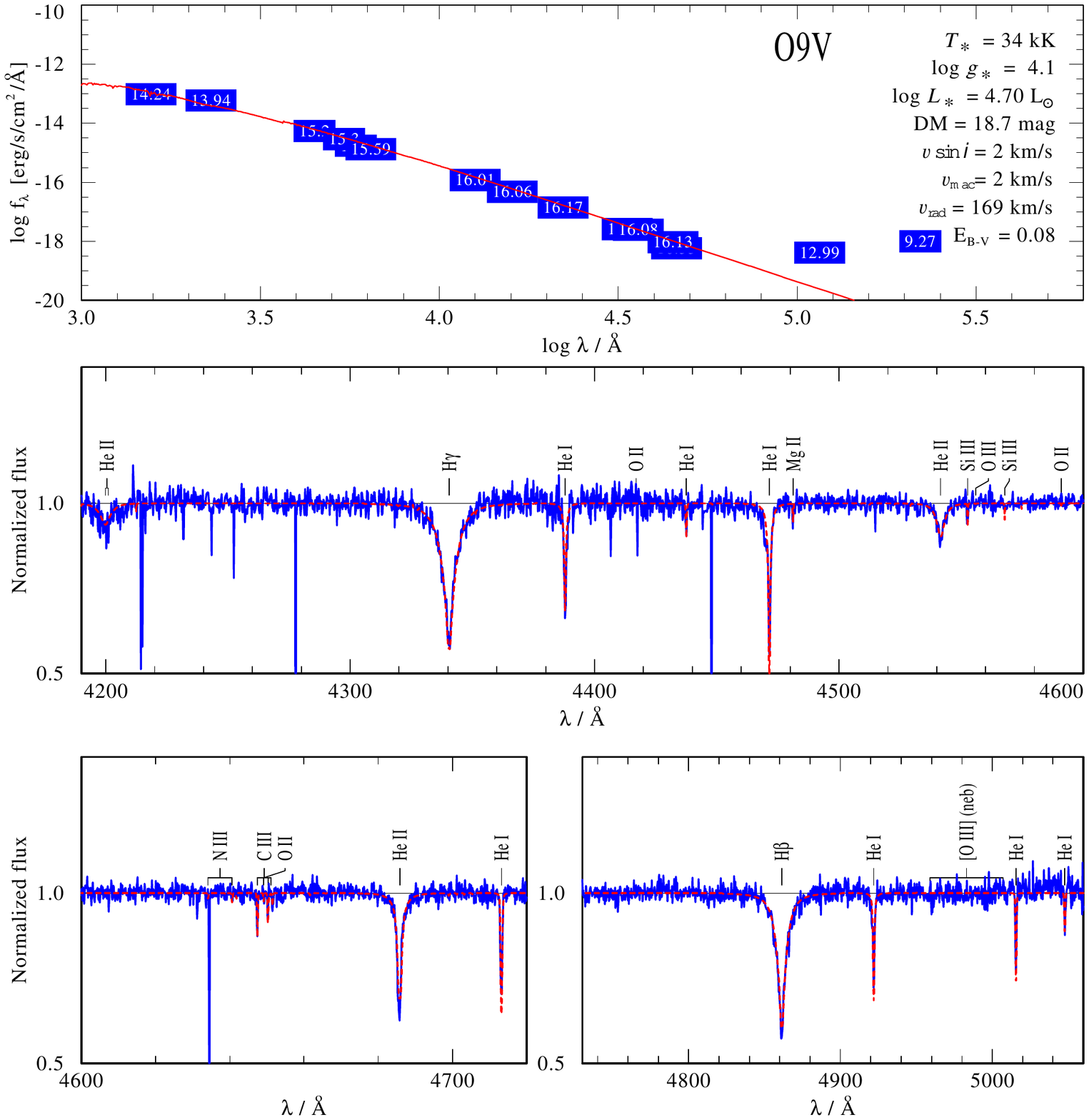}
    \caption{Same as in Fig.\,\ref{fig:MBO1} but for the star MBO2.}
    \label{fig:MBO2}
\end{figure*}

A particular PoWR model is specified by the luminosity $L$, stellar temperature $T_\ast$, surface gravity $g_\ast$\footnote{$g_\ast$=$GM_\ast/R_\ast^{2}$ refers only to gravitation; the {\em effective} gravity reduced by the outward radiation pressure might be significantly lower, depending to the star's proximity to the Eddington limit}, and mass-loss rate $ \dot{M} $ as main parameters. We constrain the stellar temperature mainly from the silicon and helium ionization balance. The surface gravity is measured by fitting the pressure-broadened wings of the Balmer lines. Typical uncertainty in $T_\ast$ and  $\log g_\ast$ are 1\,kK and 0.1\,dex respectively. For our sample stars, no UV spectra are available, while their optical spectra do not cover the H$\alpha$ line. All other lines are purely photospheric and do not show a clear wind contribution.
Hence, we are not able to empirically probe they mass-loss rate. Instead, we adopt mass-loss rates based on the $L-\dot{M}$ relation of OB stars in the SMC Wing given in \citet{Ramachandran2019}. 

We determine the luminosity $L$ and the color excess $E _{\rm B-V} $ of the individual OB stars by fitting the model spectral energy distribution (SED) to the photometry. Here, the model flux is scaled with a distance modulus of 18.7\,mag, which corresponds to the average Bridge distance \citep{mackey2017}. The uncertainty in the luminosity determination is an error propagation from color excess, temperature, and observed photometry. All these uncertainties give a final accuracy of about 0.2 in $\log L/L_{\odot}$.

The PoWR models account for complex atomic data of H, He, C, N, O, Mg, Si, P, S, and Fe group.  Initially, we calculated models with $Z=1/7\,Z_{\odot}$, based on abundances for O stars in the SMC Wing \citep{Ramachandran2019}. Subsequently, mass fractions of He, C, N, O, Si, and Mg were adjusted when necessary to best fit the observed spectra.

The analysis and derived parameters of individual objects are described below.

\begin{figure*}
    \centering
    \includegraphics[width=\textwidth,trim={0 6cm 0 0}]{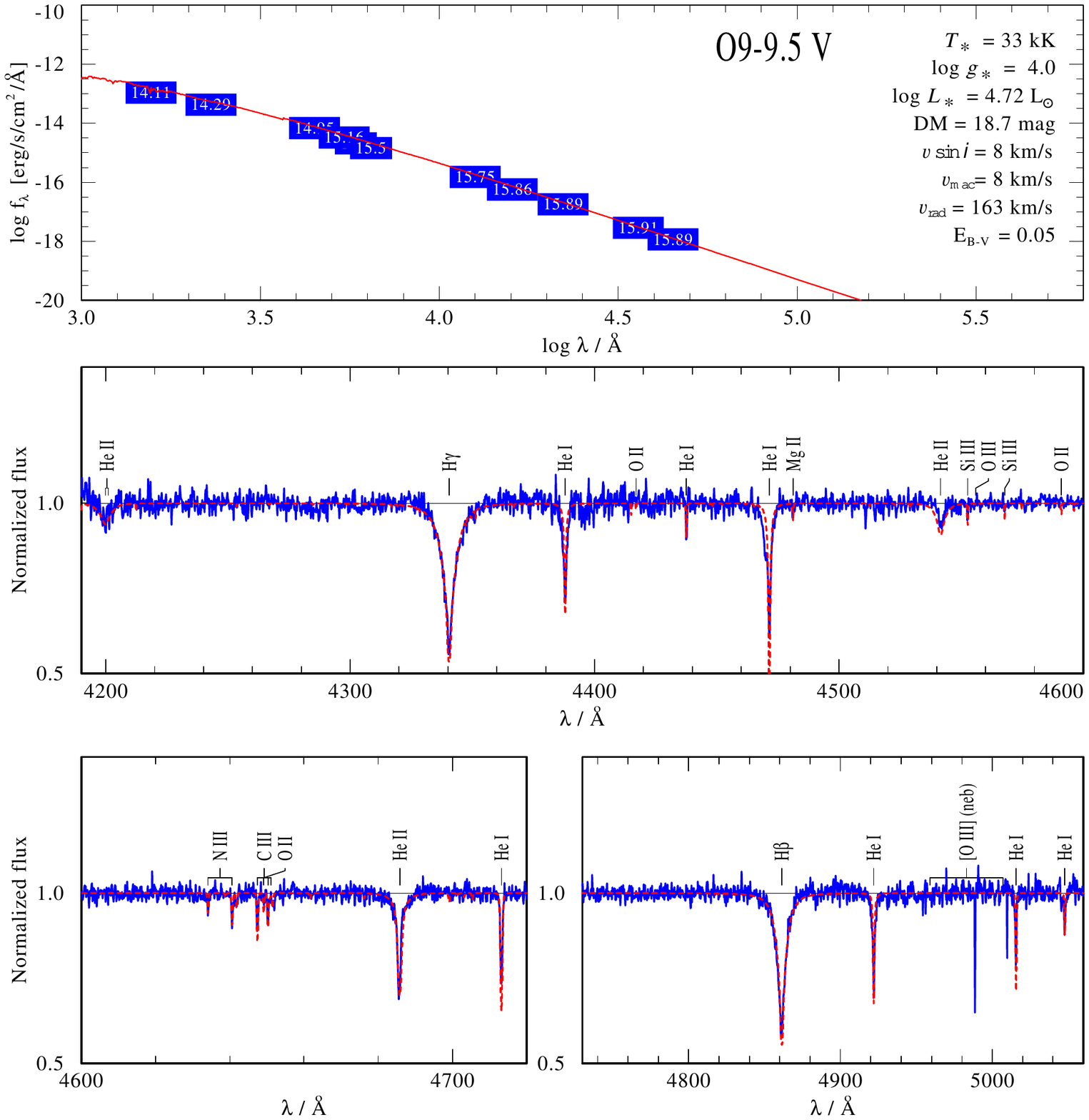}
    \caption{Same as in Fig.\,\ref{fig:MBO1} but for the star MBO3}
    \label{fig:MBO3}
\end{figure*}

\subsection{MBO1}

Although we suspect MBO1 to be a binary, the RV variations are small and the spectra do not contain lines from the secondary, suggesting the secondary could be a lower mass star. Hence for spectral analysis, we consider single star models.

Based on the strength of \hei and \heii lines, we varied the 
stellar temperature $T_\ast$ of the model from 26\,kK to 36\,kK, to achieve the best fit. Simultaneously, we varied the surface gravity $\log\,g_\ast$ from 3.0 to 4.0 to fit the Stark-broadened wings of the Balmer lines. The spectrum is best reproduced with a model of $T_\ast = 33$\,kK and $\log\,g_\ast=3.5$\,cm\,s$ ^{-2} $ (see Fig.\,\ref{fig:MBO1}). 
The line broadening also accounts for microturbulence. 
We adjusted the microturbulent velocity $\xi=12$\,km\,s$^{-1}$ which leads to a good overall fit of the metal lines.

We constrained the projected rotation velocity \vsini from the carbon and oxygen line profile shapes using the \texttt{iacob-broad} tool \citep{Simon-diaz2014}. 
We derived $\varv\,\sin i=150\pm10$\,km\,s$^{-1}$ from the Fourier Transform method and macroturbulent velocity of $\varv_{\mathrm{mac}}\sim80$\,km\,s$^{-1}$ using the goodness of fit analysis. Subsequently, convolving the model spectra with these rotational and macroturbulent velocities yields consistency with the observed spectra. 

Even though we derived the effective temperature of the model from the He ionization balance, the \hei lines at 4388\,\AA\, and 4922\,\AA\, were found to be weaker in the model. This could be only fixed by increasing the He mass fraction to $Y\sim0.35$. 
The carbon and oxygen abundance were deduced from \ion{C}{iii} and \ion{O}{ii} blend complex near 4650\,\AA. For the determination of the nitrogen abundance, we used the \ion{N}{iii} absorption lines in the wavelength range 4510--4525\,\AA\, and 4634--4640\,\AA. MBO1 shows carbon and silicon abundances $\sim -0.7$\,dex lower than solar, whereas oxygen and nitrogen are only depleted by $\sim -0.2$\,dex. This is much higher than the mean chemical abundances of the Bridge from previous studies of the ISM and B stars  \citep{Rolleston1999,lee_2005_chemical,Lehner2008}. The \ion{Mg}{ii} lines at 4481\,\AA\, were found to be much too strong for an O8 giant spectral class. We found that models adopted with an average Mg abundance of SMC, LMC, and Milky Way B stars \citep[adopted from][]{Hunter2007}  generate weaker \ion{Mg}{ii} absorption line than in the observation. We increased the magnesium abundance close to solar value in order match the observed \ion{Mg}{ii} profile as shown in Fig.\,\ref{fig:mg}. The best-fit model (red) has [Mg/H] $\sim 0.1$\,dex higher than solar.  Another possibility would be the \ion{Mg}{ii} lines comes from a cooler companion star. However, radial velocity of the line is in agreement with other lines and there are no indication of cool companion in other absorption lines.

\subsection{MBO2 and MBO3}
Figures\,\ref{fig:MBO2} and \ref{fig:MBO3} show the spectral fit for the stars MBO2 and MBO3, respectively. Spectra of MBO2 and MBO3 are similar, except that the \heii/\hei ratio is slightly higher and the metal lines are slightly weaker in MBO2. The stellar temperature of both stars was derived based on \heii/\hei ratio and \ion{Si}{iii\,$\lambda4553$}/\,\ion{He}{ii\,$\lambda4542$}. Broad wings of Balmer lines suggest that both are main-sequence stars. The most noticeable features in the spectra are the extremely narrow \hei and metal lines.

\begin{figure}
    \centering
    \includegraphics[width=0.45\textwidth,trim={1.0cm 18cm 9.5cm 2cm}]{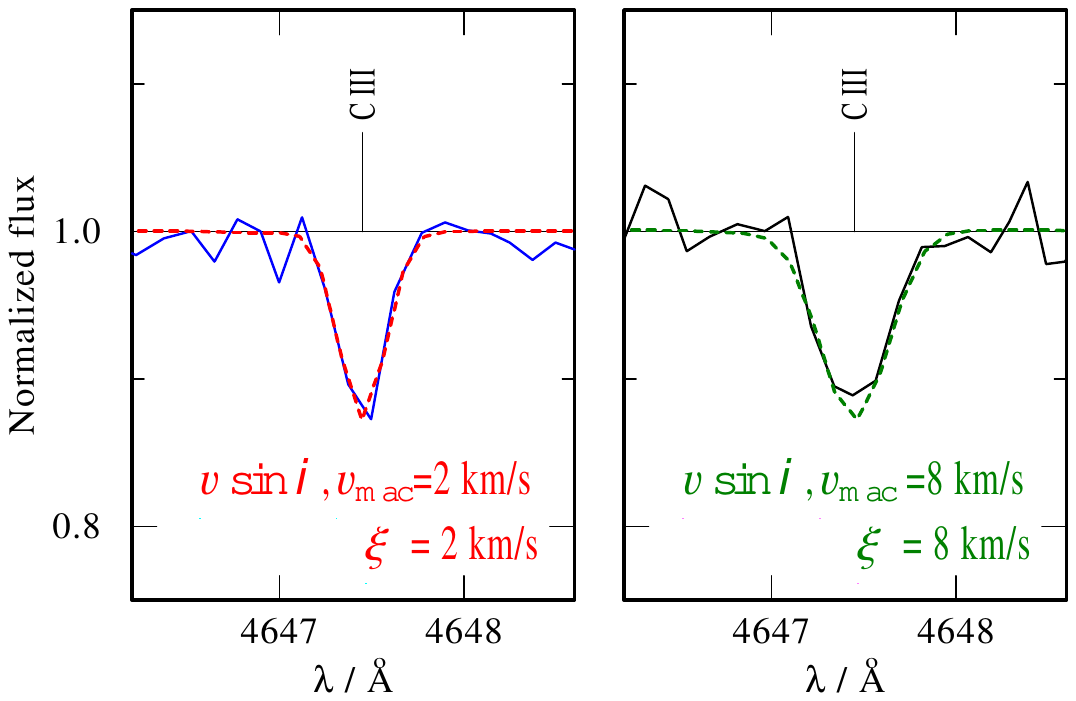}
    \includegraphics[width=0.45\textwidth,trim={1cm 18cm 9.5cm 1cm}]{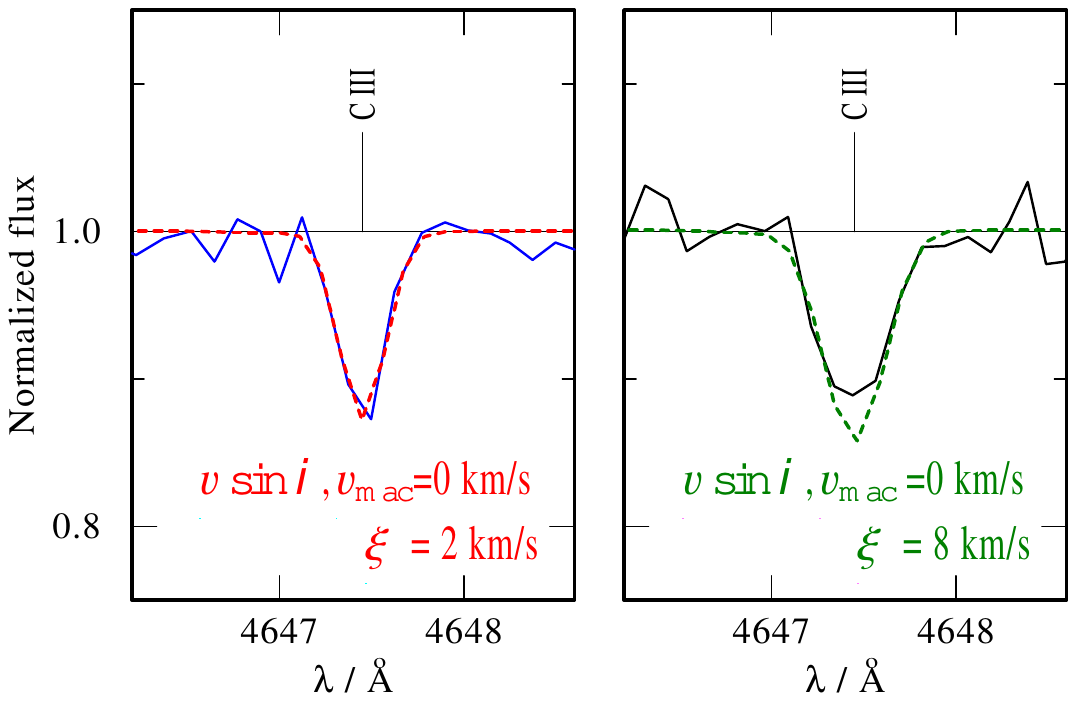}
    \caption{Left: observed \ion{C}{iii} profile of MBO2 (solid blue)  compared with best-fitting model convolved with $\xi=2$ (dotted red). Right:  observed \ion{C}{iii} profile of  MBO3 (solid black) compared with best-fitting model convolved with 8\,km\,s$^{-1}$ (dotted  green). In the lower panels, models are convolved with zero rotation, where as in the upper panels both \vsini, and $\varv_\mathrm{mac}$ are same as adopted  $\xi$.}
    \label{fig:broad}
\end{figure}

\begin{figure*}
    \centering 
    \sidecaption
    \includegraphics[width=12cm,trim={1cm 19cm 0cm 1cm}]{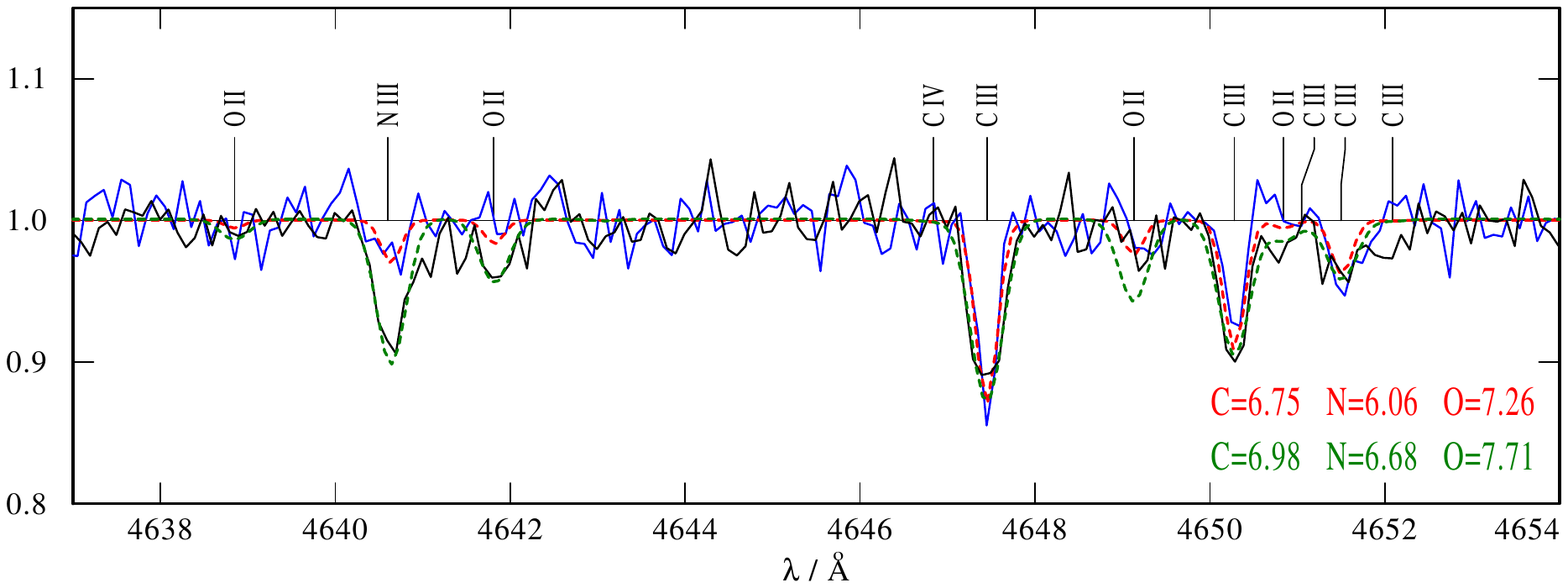}
    \caption{Comparison of \ion{C}{iii}, \ion{N}{iii} and \ion{O}{ii} lines in MBO2 (solid blue) and MBO3 (solid black) spectra. The carbon, nitrogen and oxygen abundance in the over-plotted best-fitting model of MBO2 (dotted red) is lower compared to that of MBO3 (dotted green) by a factor of two, four and three, respectively. }
    \label{fig:niii}
\end{figure*}

In the case of stars with a very slow rotation, it is difficult to constrain their \vsini using the Fourier transform method, hence we compare the broadened theoretical spectra with observation. We consider microturbulent, macroturbulent, and instrumental broadening along with rotation. 
For the smallest rotational velocities (\vsini $\lesssim$ 50\,km\,s$^{-1}$), the shape of the line is dominated by the instrumental profiles. The GIRAFFE spectra have an instrumental broadening that is well approximated by a Gaussian with an FWHM of about $0.2 $\,\AA\, for HR and about $0.6 $\,\AA\, for LR  spectra, respectively. So, the models are degraded to the instrumental spectral resolutions by convolution.

We mainly used metal lines from  \ion{C}{iii}, \ion{O}{ii}, and \ion{Si}{iii} to estimate the broadening. As an example, Fig.\,\ref{fig:broad} shows the comparison of a \ion{C}{iii} profile with our best-fitting model, convolved with corresponding values of $\xi$, \vsini, and $\varv_\mathrm{mac}$. Values of \vsini and $\varv_\mathrm{mac}$ are zero in the lower panels whereas in the upper panels they are the same as the $\xi$. A comparison of these plots suggests that microturbulence is the dominant broadening factor compared to the rotation.
In the case of MBO3 (black and green) we conclude \vsini, $\varv_\mathrm{mac} \sim 8\pm 2$\,km\,s$^{-1}$. The model spectral lines for MBO2 with a microturbulence of 2\,km\,s$^{-1}$ have a similar width as observed. Here, microturbulence imposes an upper limit below which \vsini and $\varv_\mathrm{mac}$ cannot be correctly derived. Therefore, we could only give an upper limit for \vsini and $\varv_\mathrm{mac}$. 

These main-sequence O stars with low projected rotational velocities are ideal for estimating chemical composition and thus probe the abundance of the local ISM of the Bridge. Although these stars both have very similar stellar parameters and rotate very slowly, their CNO abundances are found to differ significantly. The differential abundance analysis of MBO3 reveals a general metal depletion by  1.1\,dex relative to solar. The mean metallicity of the star is in agreement with previous abundance analyses performed for B stars and the diffuse gas in the Bridge \citep{rolleston_1999_chemical,lee_2005_chemical,dufton_2008_iron,Lehner2008}. Helium, silicon, and magnesium abundances are found to be the same in both stars. Silicon is slightly under-abundant compared to SMC stars, whereas magnesium has similar values.
On the other hand, MBO2 is highly deficient in nitrogen and oxygen compared to other Bridge stars. Figure\,\ref{fig:niii} displays \ion{C}{iii}, \ion{N}{iii} and \ion{O}{ii} lines of MBO2 and MBO3. Although $\xi$ is lower in MBO2, its \ion{N}{iii} and \ion{O}{ii} lines are extremely weak compared to MBO3. In order to fit the observation, we decreased the mass fraction of nitrogen by a factor of four, oxygen by a factor of three and carbon by a factor of two compared to MBO2. 

\subsection{Composite models}
\label{sec:composite}
Even though single star models fairly well reproduce the observed spectra of MBO2 and MBO3, the observed \hei lines are asymmetric compared to the model. Especially the blue wing of \hei lines at 4471, 4388, and 4922\,\AA\, are broader. 
We investigated the possibility that these extended wings are a weak signature of some stellar wind. However, with various test calculations in the plausible parameter range of $\log \dot{M}$=-7.5 to -9 and $\varv_\infty$=1000 to 2000\,km\,s$^{-1}$, we could not find any model which would reproduce the observed blue extensions of the \hei lines without spoiling to whole rest of the spectral fit. Therefore, we suspect that these features come from a cooler and less luminous companion. However, multi-epoch spectra of both stars don't show significant RV variations (see Sect.\,\ref{radvel}). Only the secondaries exhibit some RV dispersion. This could be explained if both stars are long-period binaries, and the primary dominates the overall observed spectra. 

\begin{figure}
    \centering
    \includegraphics[width=0.42\textwidth,trim={1cm 20cm 12cm 0cm}]{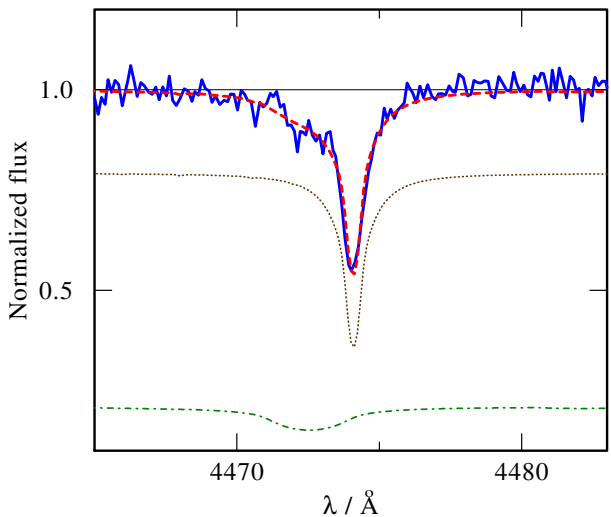}
    \caption{ \hei profile at 4471\,\AA\, of MBO2 (blue solid line).  The composite model (red dashed) is the weighted sum of O9\,V (brown dotted) and B0.5\,V (green dashed) model spectra with effective temperature 34\,kK and 29\,kK, respectively. The B star is rapidly rotating.}
    \label{fig:bin}
\end{figure}

To probe the hypothesis on binarity, we fitted the observed spectra by a composite binary model, where the primary spectrum is the same as in the single-star model fit. Figure\,\ref{fig:bin} shows the \hei profile at 4471\,\AA\, of MBO2, overplotted with our best-fit composite model. Here the radial velocity of the secondary spectrum is shifted to the blue by $\sim 60$\,km\,s$^{-1}$. The light ratios of both components are constrained from the narrow and broad absorption components. From the observed \hei profile it is clear that the secondary rotates faster  ($\sim100$\,km\,s$^{-1}$) compared to the near standstill primary O star. Such broad features are not visible in any other line. Hence, we chose the secondary model such that it does not affect \heii or metal lines in the observed spectra and only contributes to \hei lines. A reasonable fit was obtained by taking a B star model with  $T_\ast = 29$\,kK, $\log L/L _{\odot}=3.9$ and $\log\,g_\ast=4.2$ for the secondary component.

In the case of MBO3, the observed features are similar, except the secondary rotates slightly faster ($\sim140$\,km\,s$^{-1}$) and the strength of \hei absorption from the secondary is higher. A comparison of the \hei narrow and broad components suggests that the luminosity of the secondary is $\log L/L _{\odot}=4.2$.  We reproduced the observed \hei lines by assuming a B0 model with $T_\ast = 31$\,kK and $\log\,g_\ast=4.2$ for the secondary. Here also we tried hotter and cooler models for the primary, but the models we used in our single-star model fit give the best fit composite model.  Besides, the composite model better reproduced the \heii lines compared to the single star fits due to the dilution of the continuum by the B star. 

The full composite model fits for MBO2 and MBO3 are given in the Appendix\,\ref{sect:appendixa}. 
However, it should be noted that models are not a unique solution for the secondary, as cooler models also provide reasonable fits. We chose these models because the ages of the secondary (based on the position in the HRD) are in a good agreement with the primary (see Fig.\,\ref{fig:MIST_HRDbin}).

\begin{table*}
\caption{Stellar parameter} 
\label{table:stellarparameters}
\centering
\setlength{\tabcolsep}{5pt}
\begin{tabular}{ccccccccccccl}
\hline 
\hline
\noalign{\vspace{1mm}}

 Star & Spectral type &	$T _\ast$ 	& $\log\,L$ &	$\log\,g_\ast$ &$E_{\rm B-V} $&$R _\ast$& $\xi$ 	& $\varv$\,sin\,$i$&$\varv_{\rm mac}$& $\varv_{\rm rad}$&$M  _\ast$ &	$\log\,Q_{0}$ \\ 

 & & [kK] & [$L _{\odot}$]&[cm s$ ^{-2} $] &[mag]& [$R _{\odot}$] &[km\,s$^{-1}$]&[km\,s$^{-1}$]&  [km\,s$^{-1}$]&[km\,s$^{-1}$]&[$M _{\odot} $]&[s$ ^{-1} $] \\

\noalign{\vspace{1mm}}
\hline 
\noalign{\vspace{1mm}}

MBO1 	& O8\,III & 33.0 	& 5.25 	& 3.5  	& 0.14  	& 12.9 &15	& 150 &80	&  186 	& 19.3 	& 48.71  	\\ 
MBO2 &O9\,V	&34.0 	& 4.70 	& 4.1  	& 0.08  	& 6.5 	 &$\lesssim2$	& $\lesssim2$ & $\lesssim2$	& 169 	& 19.2 	& 48.00  \\ 
MBO3 &O9.5\,V	& 33.0 	& 4.72 	& 4.0  	& 0.05  	& 7.0   &$8$	& $8$ 	& $8$ 	& 163 	& 18.0 	& 47.98   \\ 
\hline
\end{tabular}
    
\end{table*}

\begin{table*}
\centering
\caption{Absolute abundances in units of $12 + \log$\,(X$_{i}/$H)}
\label{table:abundance}
\setlength{\tabcolsep}{5pt}
\begin{tabular}{cccccccccc}
\hline 
\hline
\noalign{\vspace{1mm}}

& He & N & C & O & Si & Mg & $\Sigma$CNO\\
\noalign{\vspace{1mm}}
\hline 
\noalign{\vspace{1mm}}
MBO1 & 11.12 & 7.56 & 7.72 & 8.48 & 7.34 & 7.70&8.59 \\
MBO2 & 10.95 & $\sim$6.06 & 6.75 & $<$7.26 & 6.46 & 6.75 &7.39\\
MBO3 & 10.95 & 6.68 & 6.98 & 7.71 & 6.46 & 6.75&7.81 \\
\noalign{\vspace{1mm}}
\hdashline
\noalign{\vspace{1mm}}
SMC\tablefootmark{*} & 10.95 & 7.17 (6.50) & 7.23 (7.35) & 8.06 (8.05) & 6.79 (6.80) & 6.74 (6.75) &8.17 \\
LMC\tablefootmark{*} & 10.95 & 7.54 (6.90) & 7.48 (7.75) & 8.33 (8.35) & 7.19 (7.20) & 7.06 (7.05) &8.45 \\
Galaxy\tablefootmark{*} & 10.95 & 7.59 (7.80) & 7.95 (8.40) & 8.55 (8.65) & 7.41 (7.50) & 7.32 (7.55) &8.68\\
\hline
\end{tabular}
\tablefoot{
\tablefoottext{*}{average (and baseline) abundance of SMC, LMC, and Galactic B stars adopted from \citet{Hunter2007}}
}
\end{table*}

\begin{table}
\centering
\caption{Abundances relative to solar values adopted from \citet{asplund2009} in logarithmic scale}
\label{table:abundancediff}
\setlength{\tabcolsep}{4pt}
\begin{tabular}{ccccccccc}
\hline 
\hline
\noalign{\vspace{1mm}}

& He & N & C & O & Si & Mg& Mean \\
\noalign{\vspace{1mm}}
\hline 
\noalign{\vspace{1mm}}
MBO1 & 0.19 & -0.27 & -0.71 & -0.21 & -0.17 & 0.10 & -0.25  \\
MBO2 & 0.02 & $\sim$-1.77 & -1.68 & $<$-1.43 & -1.05 & -0.85 & -1.36  \\
MBO3 & 0.02 & -1.15 & -1.45 & -0.98 & -1.05 & -0.85 & -1.10 \\
\hline
\end{tabular}
\end{table}

\section{Results and discussions}
\label{sec:results}

The stellar parameters obtained by our spectral analysis of the three Bridge O stars are given in Table\,\ref{table:stellarparameters}, while in Table\,\ref{table:abundance} the chemical abundances compared to average values for SMC, LMC, and Galactic B stars. Abundances relative to solar and the average metal depletion of the stars are given in Table\,\ref{table:abundancediff}.  The derived abundances and parameters are based on the single star model fits. The binary analysis of MBO2 and MBO3 leads to a reduction of the luminosity by $\sim 0.1$ dex and the stellar mass by a factor of $\sim 3-4$. Since the primary O star dominates the observed spectra, the rest of the derived values are nearly unaffected.

Uncertainties of the abundances we estimate to 0.1–0.2\,dex, typically. This excludes
uncertainties arising from observational errors and from the other model parameters. Due to stronger rotational broadening for MBO1, these uncertainties are higher than for MBO2 and MBO3.
In case of MBO2, the oxygen lines are very weak and it is difficult to distinguish them from noise given the S/N of the spectra. So we only provide an upper-limit for the oxygen abundance in MBO2. The measurement error for Mg is higher, since the abundance is derived from only one line, while N and Si abundances are measured from two lines, and C, O and He abundances are averaged from multiple lines. However, the Si abundance is less accurate because Si lines are more susceptible to uncertainties in microturbulence and stellar temperature.

In Fig.\,\ref{fig:MIST_HRD} we locate the three Bridge O-stars in the Hertzsprung–Russell diagram (HRD) and compare with stellar evolutionary tracks and isochrones from the MIST series \citep[MESA Isochrones and Stellar Tracks,][]{Choi2016} corresponding to [Fe/H]\,=\,-1. Both tracks and isochrones assume an initial rotation of $\varv/\varv_\mathrm{{critical}}\,=0.4$. We used the MIST web-intepolator\footnote{http://waps.cfa.harvard.edu/MIST/} to find the best-matching tracks and isochrones for the analyzed O stars and plotted them in Fig.\,\ref{fig:MIST_HRD}.  Since we suspect that MBO2 and MBO3 are binaries, positions of the primary and secondary derived from the composite spectral analysis (see Sect.\,\ref{sec:composite}) are shown in Fig.\,\ref{fig:MIST_HRDbin}.
The estimated masses and ages of the MBO stars are tabulated in Table.\,\ref{tab:agemass}. The evolutionary age and mass of the primaries are found to be similar to that derived from single star analysis within the uncertainty limits. 

The inferred ages of these newly discovered O stars give evidence for massive star formation in the Bridge as recent as within the last 7-8\,Myr. This is lower than previously suggested values \citep{Demers1998,Irwin1990} that assume the population consisting only of B or later-type stars.  The star formation history derived by \citet{Harris2007} suggested two distinct recent episodes $\sim 160$ and 40\,Myr ago. Interestingly, no massive YSOs ($M > 10M_{\odot}$) were found in the Bridge so far \citep{Chen2014}. The discovery of O stars gives direct evidence for massive star formation continuing in the tidal Bridge, even a few hundred Myr after the encounter between the Clouds. 

Based on atmospheric parameters, abundances, and positions in the HRD, the characteristics of these O stars along with possible explanations are discussed below. 

\subsection{MBO1 -- a metal-rich O giant in the Bridge}  

MBO1 is one of the most luminous and massive $M\sim25\,M_{\odot}$ star in the whole Bridge. The chemical abundances in MBO1 are strikingly different from the other two O stars in our sample, and rather comparable to LMC and Galactic O stars. The Bridge star DI\,1388 analyzed by \citet{Hambly1994} also has higher metal abundances than the average Bridge value, similar to MBO1. However, ISM gas-phase abundances derived by \citet{Lehner2008} from UV observations of the same star are in agreement with a metal depletion by -1\,dex.

Interestingly, the magnesium abundance in MBO01 is nearly solar, much higher than other elements. This also true for other two sample stars when comparing their Mg abundance to other elements (see Table.\,\ref{table:abundancediff}).  The overabundance of Mg relative to Fe is usually found in metal-poor or mildly metal-poor stars (intermediate or low mass) in the disk and halo of the Milky Way \citep[e.g.,][]{Bensby2014,Argast2002, Abia2004,Cayrel2004}.  However, only thin disk stars in the Milky Way show  [Mg/H]>0 and typically with [Fe/H]>0, where as thick and halo populations usually have [Mg/H] < 0.

\begin{figure}
    \centering
    \includegraphics[width=0.45\textwidth]{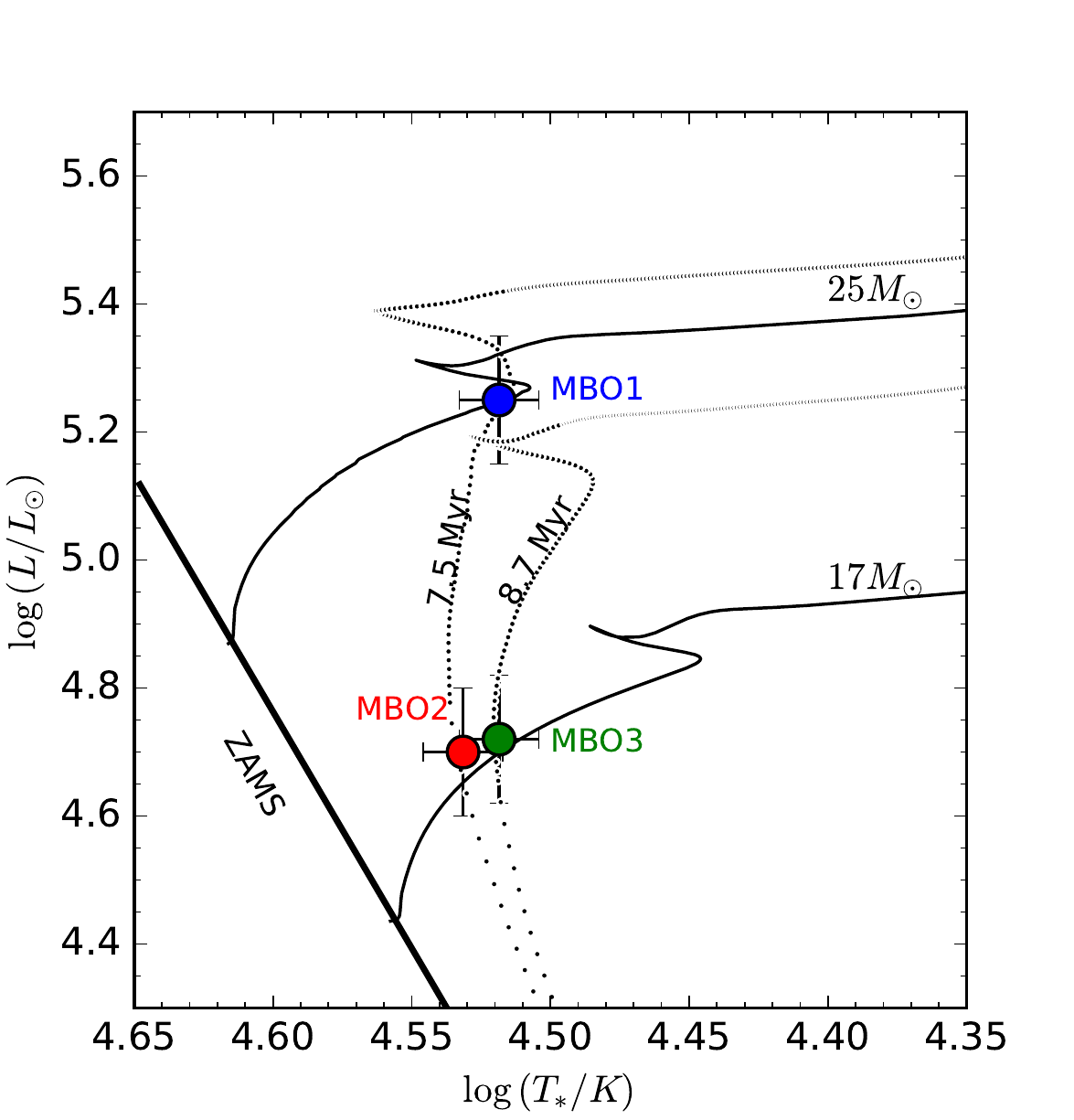}
    \caption{HRD for Bridge O stars. MIST tracks (solid) and isochrones (dotted) for [Fe/H]=-1 and initial rotational velocity of $\mathrm{\varv_{ini} =0.4\varv_{critical}}$ \citep{Choi2016} are shown for comparison. }
    \label{fig:MIST_HRD}
\end{figure}

\begin{figure}
    \centering
    \includegraphics[width=0.45\textwidth]{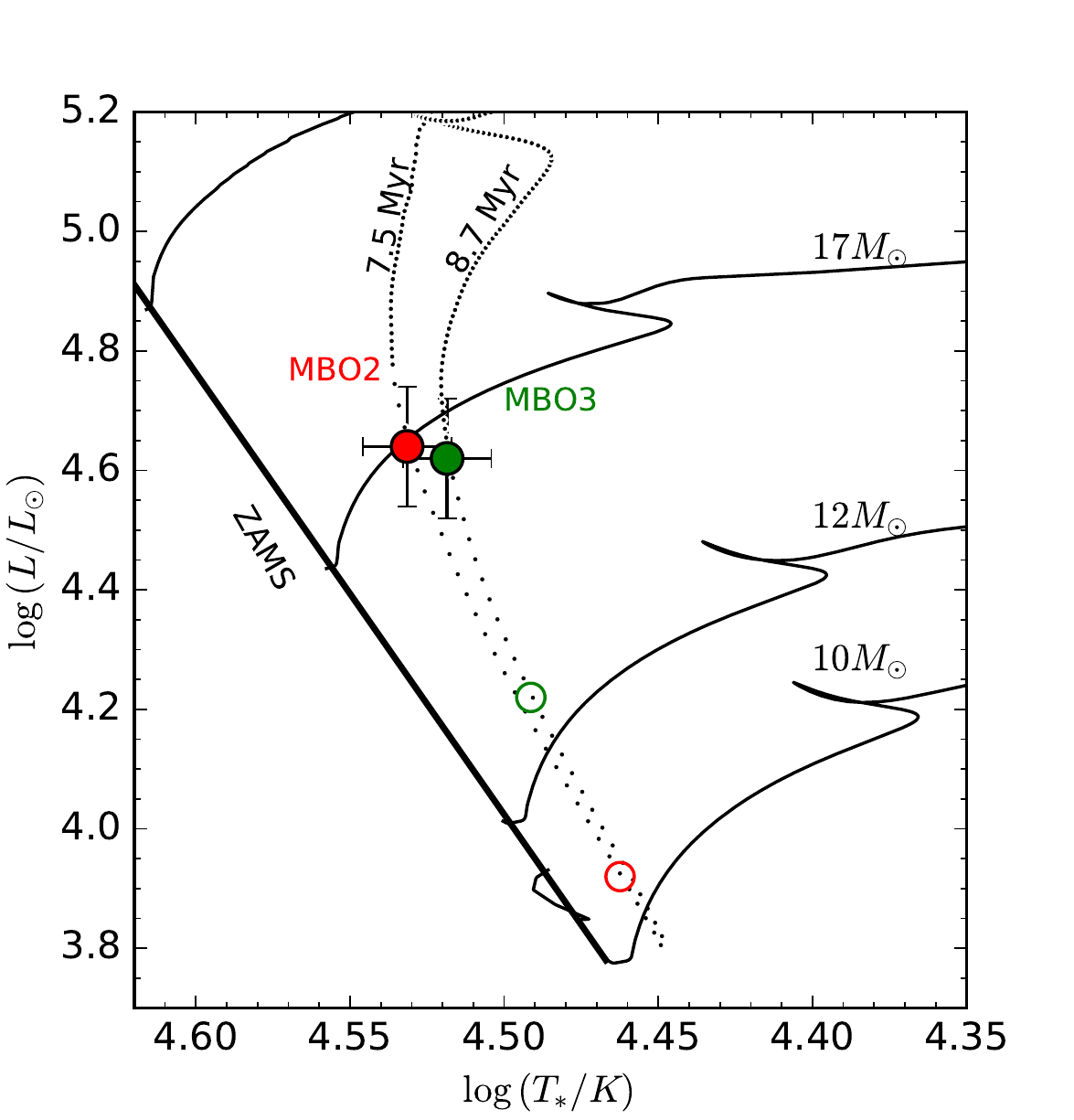}
    \caption{Same as Fig.\,\ref{fig:MIST_HRD} but positions derived from binary analysis of MBO2 and MBO3. The primary and secondary are marked with filled and open circles, respectively. }
    \label{fig:MIST_HRDbin}
\end{figure}

Figure\,\ref{fig:MIST_HRD} shows that MBO1 nearly reached the main sequence turn-off.
Nuclear products like helium and nitrogen are expected at the surface of the star, based on its position in the HRD and its fast rotation. In such cases, one would expect the depletion of carbon and oxygen to balance the CNO equilibrium.  However, all the heavy elements have consistently higher mass fractions than other Bridge or even SMC stars. This is not consistent with CNO nucleosynthesis or chemical mixing.

Both the radial velocity and Gaia proper motions of MBO1 are in good agreement with the Bridge population. Moreover, the radial velocity of the star matches with that of nearby \hi shells \citep{Muller2003}.
So it is very unlikely to be a Galactic foreground object. With its very young age, it would be difficult to assume this star runs away from the LMC. Given the close agreement in age with the other two nearby O stars, all of them most likely formed in the same star formation episode in the Bridge.

One could think of possible explanations for the chemical peculiarity of MBO1 due to close-binary interactions, merging, or chemical mixing due to homogeneous evolution. 
Our SED fit to the photometry (from UV to mid-IR) suggests a relatively high reddening for a Bridge star. Moreover, from $\gtrsim8\mu $m the star shows an infrared excess, suggesting the presence of circumstellar matter. This could be an indication of a previous binary interaction. The evolutionary mass of MBO1 is higher than its spectroscopic mass, i.e., the luminosity of the star is $\sim0.4$\,dex higher for its spectroscopic mass. This could be a result of helium surface enrichment due to enhanced rotational mixing. The star might have been born as a rapid rotator,  which is expected for low-Z stars and follows a quasi chemically homogeneous evolution. Fast rotation can also arise from gaining angular momentum during binary interactions. In this case, the secondary (mass gainer) would be fast-rotating, while the primary turns into a stripped He star. The relatively weak H$\gamma$ and H$\beta$ lines in the spectra could be another evidence for disk emission. Besides, RV dispersion also suggests a lower mass companion. Therefore, an Oe star with a hidden stripped companion could be one possible explanation for MBO1's peculiar nature.   

But even with this post binary interaction scenario, it is still difficult to explain the MBO1 abundances. Surface nitrogen of the core hydrogen burning mass gainers can be enhanced by factors up to $3-6$ \citep{Langer2012}. These mass gainers also spun-up sufficiently, which leads to efficient mixing and nitrogen enhancement by up to another factor of three. However, in MBO1, along with nitrogen, other metals are also enhanced by an order of magnitude. Moreover, N/O and N/C ratios are found to be small ($<1$), indicating that nitrogen is under-abundant relative to carbon and oxygen. Despite these low ratios, He is relatively enriched in the star. 

An alternative scenario could be that MBO1 is a merger product.
Surface abundance predictions for merger products are still unclear. Rotational mixing can happen in the initial stages, where nitrogen and helium are expected to be enriched. Current models do not predict the enrichment of all metals as a result of binary interactions or mixing. We conclude that the reason for the chemical peculiarly of MBO1 remains unclear. 

As another explanation for the chemical peculiarity of MBO1,  the material from which the star was formed was more metal-rich than average in the Bridge.  Recent kinematic studies and simulations \citep{Zivick2019,Oey2018} support a direct collision scenario between the LMC and the SMC. In this case, the material from which MBO1 was formed could have been accreted from the LMC. 

\begin{table}
\centering
\caption{Ages and masses of MBO stars derived from evolutionary tracks and isochrones. Spectroscopic
masses, calculated from $\log\,g_\ast$ and $R_\ast$ are given for comparison. For MBO2 and MBO3, parameters  based on single star and binary analysis are given in the first and second row, respectively.}
\label{tab:agemass}
\setlength{\tabcolsep}{6pt}
\begin{tabular}{llll}
\hline 
\hline
\noalign{\vspace{1mm}}
 & Age & $M_{\mathrm{ev}}$ &$M_\mathrm{{spec}}$  \\ \noalign{\vspace{1mm}}
 & [Myr]& [$M _{\odot} $] & [$M _{\odot} $]\\
\hline 
\noalign{\vspace{1mm}}
MBO1 &   7.5 & 25.2 & 19.3 \\
MBO2 &   7.5 & 17.5 & 19.2   \\
& 7.5 & 17 & 16 \\
MBO3 &   8.7 & 17.3 & 18    \\ 
 & 8 & 16.5 & 14\\
\hline
\end{tabular}
\end{table}


\subsection{MBO2 and MBO3 -- slow rotators at low Z}

At low metallicity, massive stars might stay hot and compact during most of their lives having weak stellar winds \citep{Ramachandran2019}. With their low metal abundance, the Bridge stars should be rapid rotators, especially on the main sequence. However, both MBO2 and MBO3 rotate very slowly ($<10$\,km\,s$^{-1}$). These stars are still on the main sequence and are unevolved based on their chemical abundances. Both stars might be coincidentally seen nearly pole-on. 
\citet{Hunter2005} discussed two sharp-lined B stars in the SMC with very low projected rotational velocities, i.e., similar to our sample stars. These B stars are also found to be  unevolved since their abundances are representative of the current baseline chemical composition of the ISM in the SMC. However, it should be noted that \citet{Hunter2005}  pre-selected these narrow-line stars from previous observations of a large sample.
Recently an excess number of slow rotating (\vsini$\lesssim40$\,km\,s$^{-1}$) apparently single B-type stars were reported in the SMC cluster NGC346  \citep{Dufton2019,Dufton2020}.
Yet, most of them are nitrogen-enriched compared to the baseline SMC value. 
Furthermore, our sample O stars are much more metal-poor than the SMC stars, nevertheless have even lower rotation rates compared to the latter.

Several processes can slow down the rotation of massive main-sequence stars, including braking due to magnetic fields \citep{Meynet2011,Morel2012,Keszthelyi2019} or stellar mergers with subsequent magnetic braking \citep{Schneider2016,Schneider2019}, tidal interactions in binaries, et cetera. 
We suspect that these narrow-lined stars are binary systems with synchronized orbits. The tidal interactions between the components may slow down the stars \citep{dufton2013}. This scenario seems to be more in line with  MBO2 and MBO3 since we already suspect them to be binaries based on features in the \hei profiles (see Sect.\,\ref{sec:composite}).

\subsection{MBO2 -- an extremely nitrogen-poor O star}

Stars MBO2 and MBO3 have similar stellar parameters and evolutionary stages. The projected spatial distance between these two stars is only 15\,pc. The radial velocities of both stars are similar and match with the adjacent \hi shell. Despite these similarities, MBO2 is strongly depleted of heavy elements.  With the depletion of nitrogen by -1.8\,dex relative to solar, MBO2 is the nearest metal-poor massive star ever analyzed. The mean metal abundance of the star is only $\lesssim 1/22\,Z_{\odot}$. 
 
Since the star lies on the main sequence, its chemical composition should reliably reflect that of the interstellar material from which it was formed. 
The much higher metal abundance of the nearby MBO3 could be an
indication of unmixed ISM in the Bridge. Ionized gas analysis by \citet{Lehner2008} found a depletion of nitrogen by -1.1\,dex toward one sightline, while it was -1.75\,dex toward another. A low nitrogen abundance (also overall metallicity) was found at the eastern end of the Bridge near the LMC, i.e., in the region far away from MBO2.

\begin{figure}
    \centering
    \includegraphics[width=0.5\textwidth]{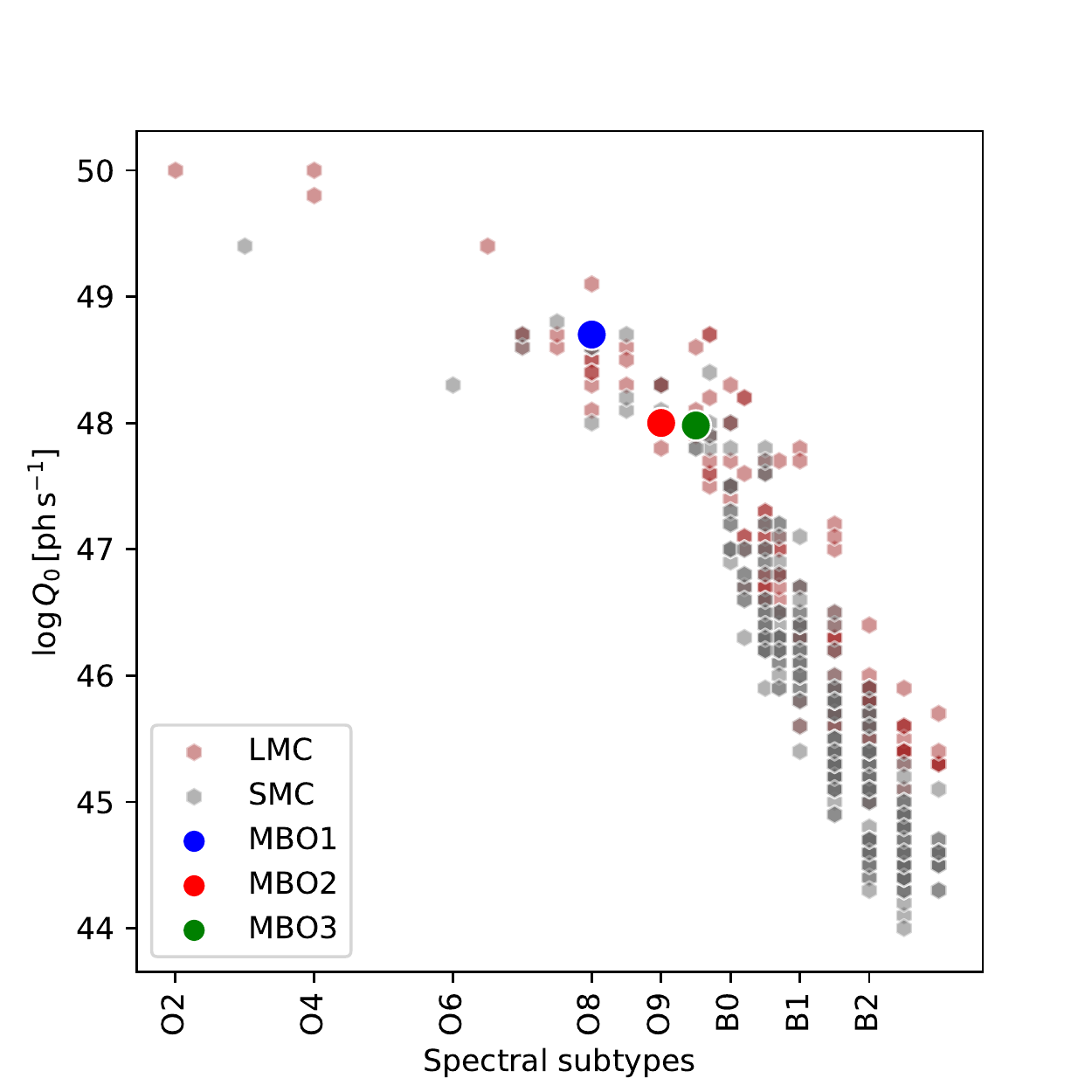}
    \caption{Ionizing photon flux vs spectral subtype for analyzed bridge O stars, comparing with OB stars in the LMC \citep{Ramachandran2018} and the SMC \citep{Ramachandran2019}.  }
    \label{fig:logQ}
\end{figure}

\subsection{Ionizing feedback}
Our sample O stars are the dominant ionizing sources of the \hii regions DEM\,169 and 168 in the western Magellanic Bridge. Using the predicted  rate of hydrogen ionizing photons ($Q_{0}$) for  each O star in our sample (see Table\,\ref{table:stellarparameters}), we estimate their combined ionizing photon flux  as  $\sim 7\times 10^{48}$\,ph\,s$^{-1}$. About 70\,\% of these photons come from the luminous O giant MBO1. In Fig.\,\ref{fig:logQ}  we compared the ionizing photon contribution of our MBO stars with the LMC and the SMC OB stars \citep{Ramachandran2018,Ramachandran2019}. The diagram showcases the relation between spectral subtypes and $\log Q_{0}$ values for different metallicities. Our analysis does not suggest higher ionizing photon flux from stars at low metallicity. Even the two metal-poor O stars in the Bridge produce a similar amount of ionizing photons as stars of the Magellanic Clouds with the same spectral types. The $\log Q_{0}$ values drop rapidly for spectral types later than B0. So, the ionizing feedback from few O stars will dominate over the B star population across the whole Bridge.

Albeit the number of ionizing photons produced by the Bridge is not very different from that of stars in the Magellanic Clouds, the size of their surrounding \hii regions will be different because of the low ISM density in the Bridge.
The overall density in the Bridge is much lower  \citep[$<0.03-0.1$\,cm$^{-3}$,][]{Lehner2001,Lehner2008} than  typical densities in  Magellanic Clouds star-forming regions   \citep[$\sim 10-100$\,cm$^{-3}$,][]{CarlosReyes2015, Wilcots1994}. Using the classical Stromgren sphere equation, the initial radius of the \hii region would reach around 100-200\,pc.  Assuming a continuous supply of photons during the evolution, the final radius of the \hii sphere around the sample O stars can reach up to 300\,pc. On the other hand,  O stars of the same spectral subtypes in the Magellanic clouds could only produce \hii regions with less than 50\,pc radius, adopting a density of 10\,cm$^{-3}$.
While the \hii regions in the Bridge are large because of
the low density, the \halpha emission (which depends on the square of the density) is faint as observe by \citet{muller_2007_h}.

\section{Implications for the Magellanic Bridge}
\label{sec:impli}
\subsection{Metallicity of the Bridge}

\begin{figure}
    \centering
    \includegraphics[width=0.5\textwidth]{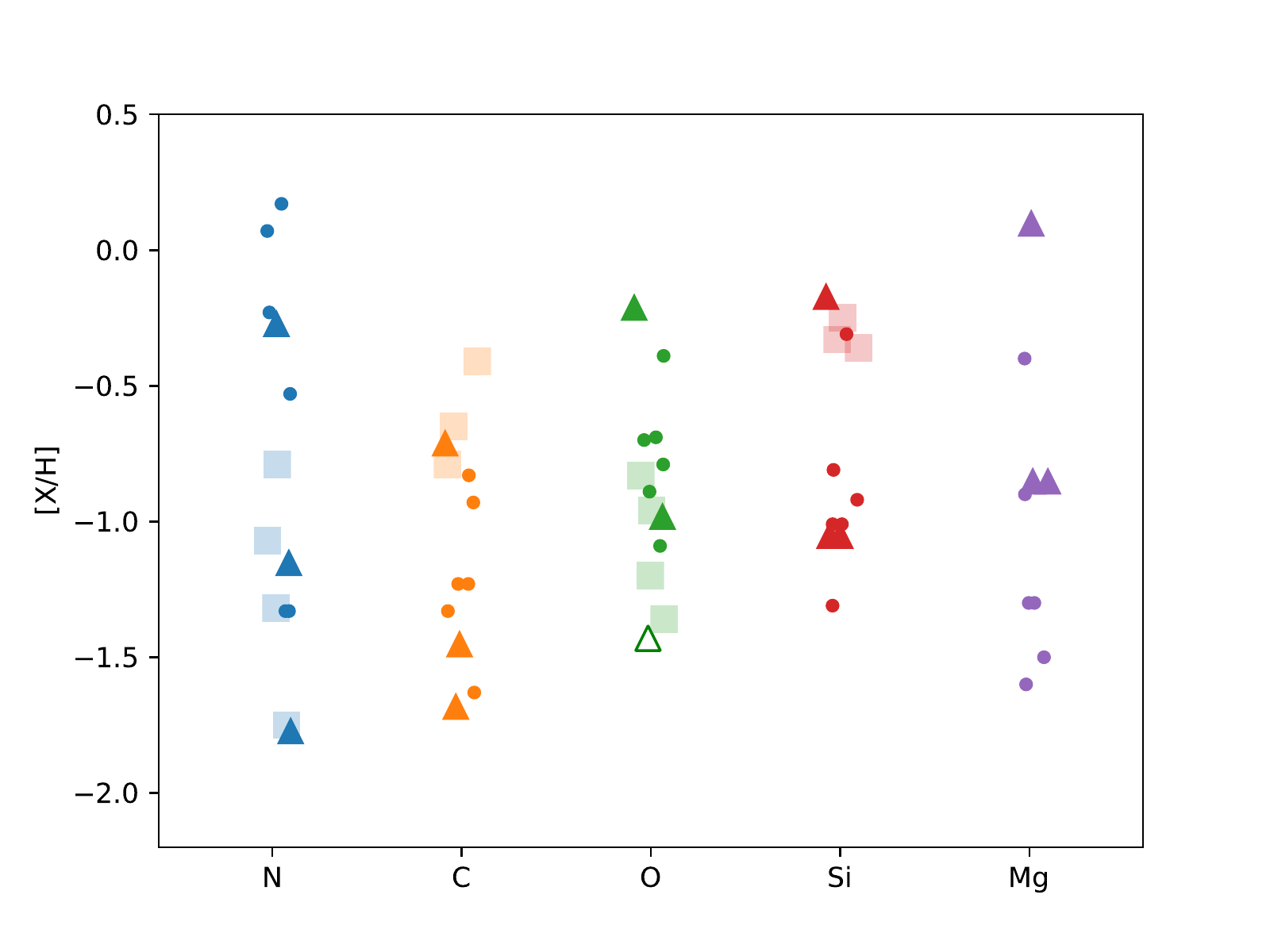}
    \includegraphics[width=0.5\textwidth]{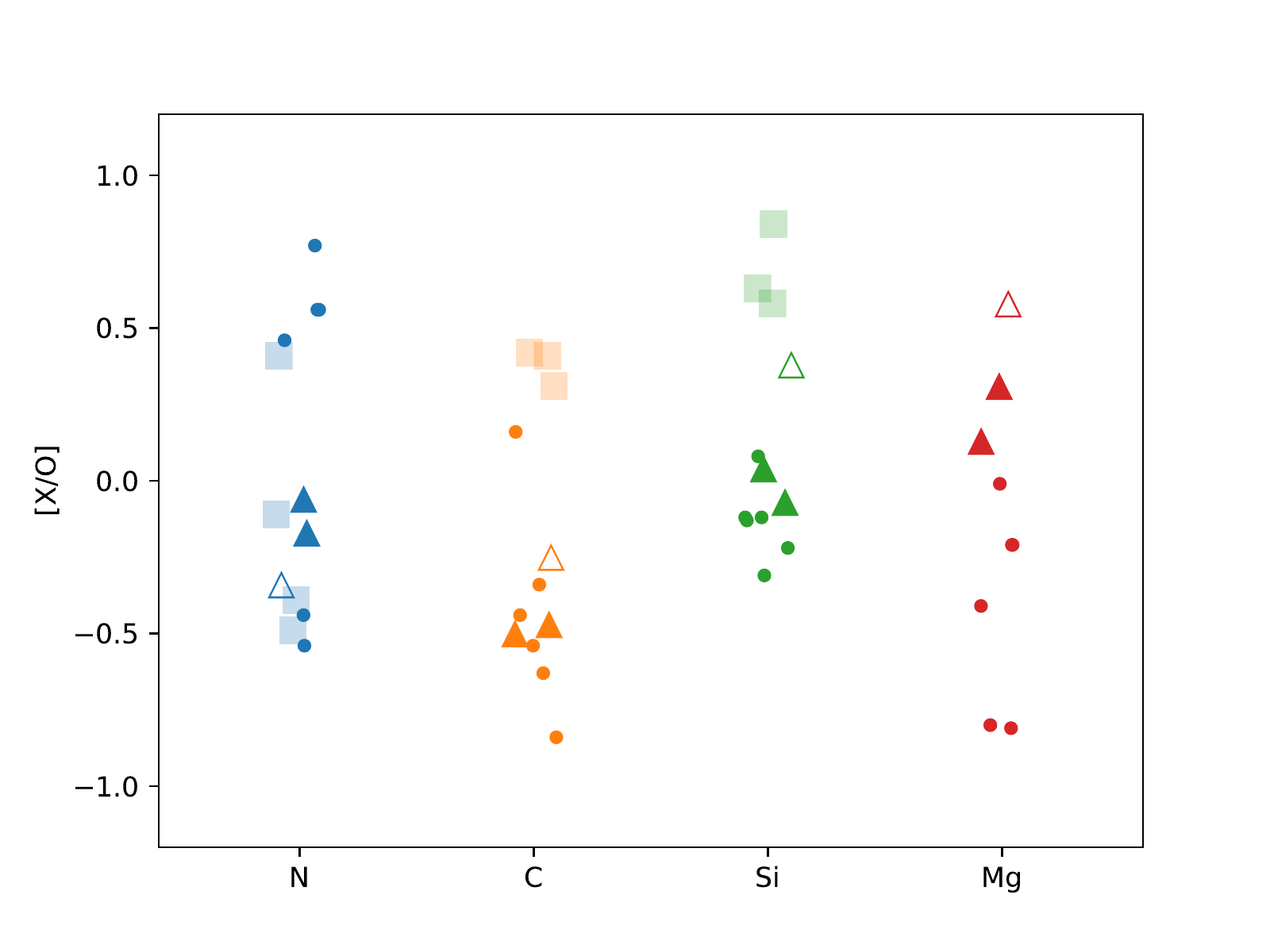}
    \caption{Depletion of metals in the Bridge relative to hydrogen (upper panel) and oxygen (lower panel). Our sample stars are marked with triangles. Open triangle refers to the upper-limit on the oxygen abundance for MBO2. For comparison, we include abundances of previously analyzed Bridge stars  \citep{rolleston_1999_chemical,lee_2005_chemical,Hambly1994} as circles, and ISM abundances \citep{Lehner2008,Lehner2001} as squares.} 
    \label{fig:abn}
\end{figure}

Although our sample O stars are located in the same star-forming region of the Bridge, their chemical abundances are found to be different. 
Figure\,\ref{fig:abn} shows the large scatter of the (relative) metal abundances of Bridge stars. The upper panel of the figure displays elemental abundances relative to hydrogen while  in the lower-panel abundances are relative to oxygen.
We also incorporated abundances of B stars as well as the ISM from previous studies. Individual surface abundance can vary from their baseline values due to nuclear burning. However, the sum of the CNO abundances is also significantly different for our sample stars (see Table.\,\ref{table:abundance}). The median values of each element in our sample stars are in agreement with the previously suggested value of $\sim-1$\,dex. However, the dispersion is also about 1\,dex for each element.

Highly scattered stellar abundance patterns of low mass dwarf galaxies and metal-poor stars in the halo have been attributed to inhomogeneous mixing and stochastic star formation effects 
\citep{James2020,Mashonkina2017,Simon2015,Cescutti2010,Cescutti2008}.
Since the winds of massive stars are weak at low Z and star formation in the Bridge was only active during the last $\sim200$\,Myr, only a few supernovae must have contributed to the chemical enrichment of the ISM. One can speculate that MBO2 originated in a region where the abundances reflect either poor mixing of metals or a lack of supernovae, whereas at the location of MBO1 the gas was enriched by a small number of earlier supernovae. This is in agreement with the conclusions of \citet{muller_2007_h} using the \halpha emission study.  They suggest that the Bridge ISM is not substantially mixed by energetic processes such as stellar winds and supernovae based on the low emission measures.

The chemical inhomogeneities could also originate from the accretion of metal-poor gas due to galaxy interactions. For example, \citet{Olsen2011} found a distinct population of low metallicity stars in the LMC and suggested that they were accreted from the SMC. 
The interaction between the LMC and SMC is going on over the last few Gyr and is likely responsible for the various star-forming episodes and the creation of the Magellanic Bridge connecting them. Dynamical simulations \citep{Gardiner1996,Besla2012} suggest that the last direct interaction between both dwarf galaxies occurred about 200\,Myr ago. More recent results claim this interaction happened only 150\,Myr ago \citep{Zivick2019}.
The abundance patterns of Bridge stars allows us to place constraints on its formation history.

If the Bridge of \hi gas connecting the Magellanic Clouds formed from stripped SMC gas during the latest encounter, chemical abundances of the young Bridge population must be compatible with that of the SMC. However, the star-to-star abundance scatter is high in the Bridge.  E.g., MBO1 is found to be quite metal-rich while MBO2 is significantly metal-poor. The transfer of metal-rich gas from the LMC to the Bridge by recent interaction can explain the higher metal abundance of MBO1. On the other hand, the low abundances of the star MBO2 can serve as an upper limit for the primordial abundance of the Bridge. The material from which the star has formed must have got accreted in the past, during the earliest stages of interactions, when the SMC had a lower metallicity. On this basis, we can constrain the time of the Bridge formation. 
Comparison with the age-metallicity relation of the SMC stars \citep{Cignoni2013,Piatti2011} suggests an age of more than 5\,Gyr corresponding to an iron depletion of at least $-1.3$\,dex. The iron abundance of the Bridge B stars derived by \citet{dufton_2008_iron} is much lower by $\sim 0.5$\,dex than their corresponding lighter element abundances. So we expect a very low [Fe/H]<-1.3 for MBO2.  
Hence we speculate that the formation of the Bridge might have been initiated a long time before the most recent encounter. The material that was stripped from the Clouds to form the Bridge must be nearly pristine gas. This is plausible since the LMC–SMC pair likely had been bound to each other for more than several Gyr and possibly even for a Hubble time \citep{Kallivayalil2013,Besla2007}.

Different stages of the gas accretion process and tidal interactions might have made the Bridge ISM chemically inhomogeneous. Since the Bride mostly consists of young stars ($<200$\,Myr), we can expect that star formation was not active before the last interaction event. Quiescent star formation for a long time implies a lack of supernovae in the ISM for chemical mixing. Therefore the Bridge might have maintained the level of abundance variation throughout history until star formation happened in the past hundred million years. 


\subsection{Ionization sources of the Bridge}
\label{sec:bridgeionization}
The Bridge contains a substantial amount of ionized gas. The main photo-ionizing agents are expected to be hot stars in the Bridge and escaping photons from Magellanic Clouds. We have estimated the ionizing feedback from the three Bridge O stars and found that their $\log Q_{0}$ follows a similar relation with spectral subtype as for LMC and SMC OB stars. So, a rough estimate of the number of O stars in the Bridge can provide their overall feedback.

\begin{figure}
    \centering
    \includegraphics[width=0.45\textwidth]{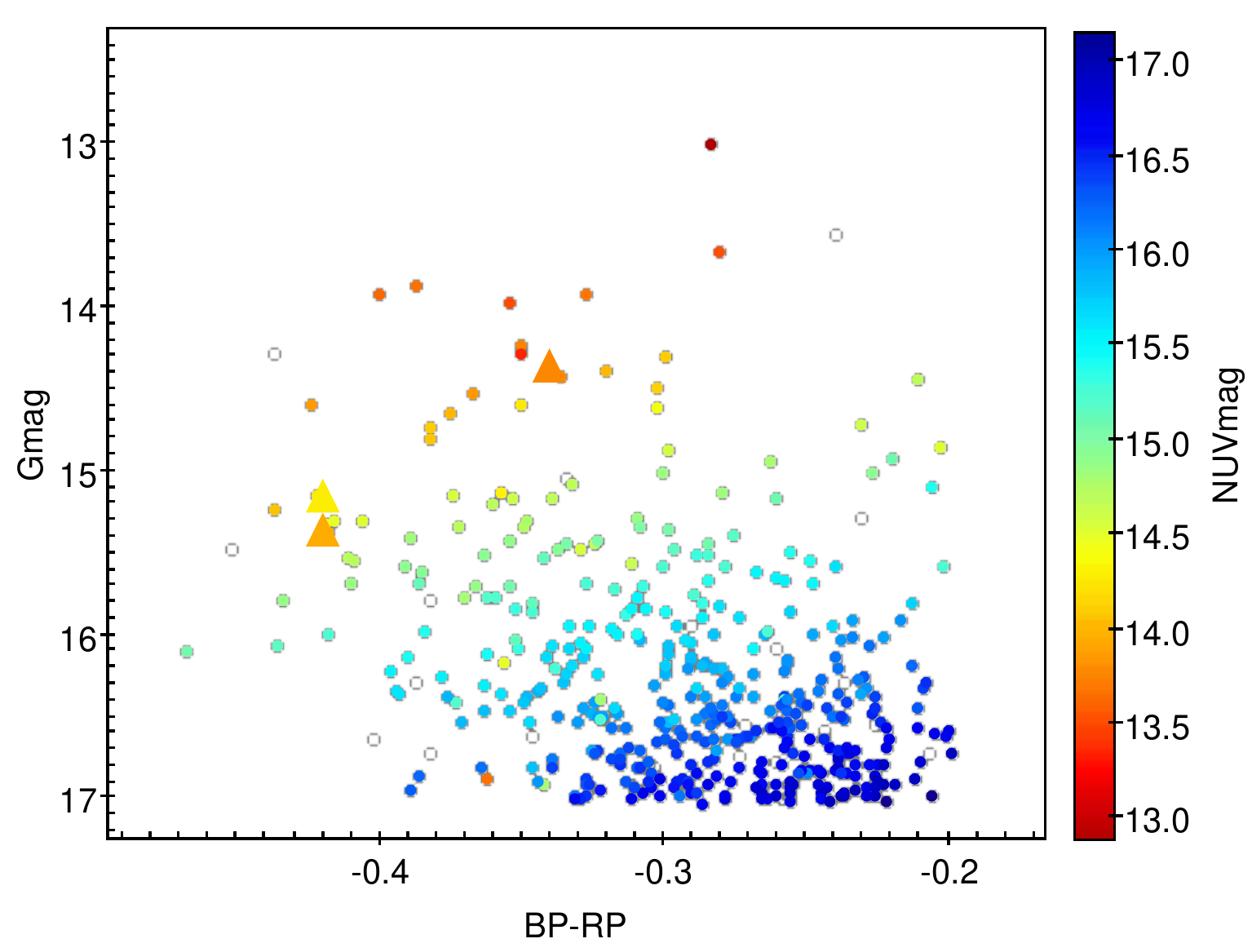}
    \includegraphics[width=0.4\textwidth]{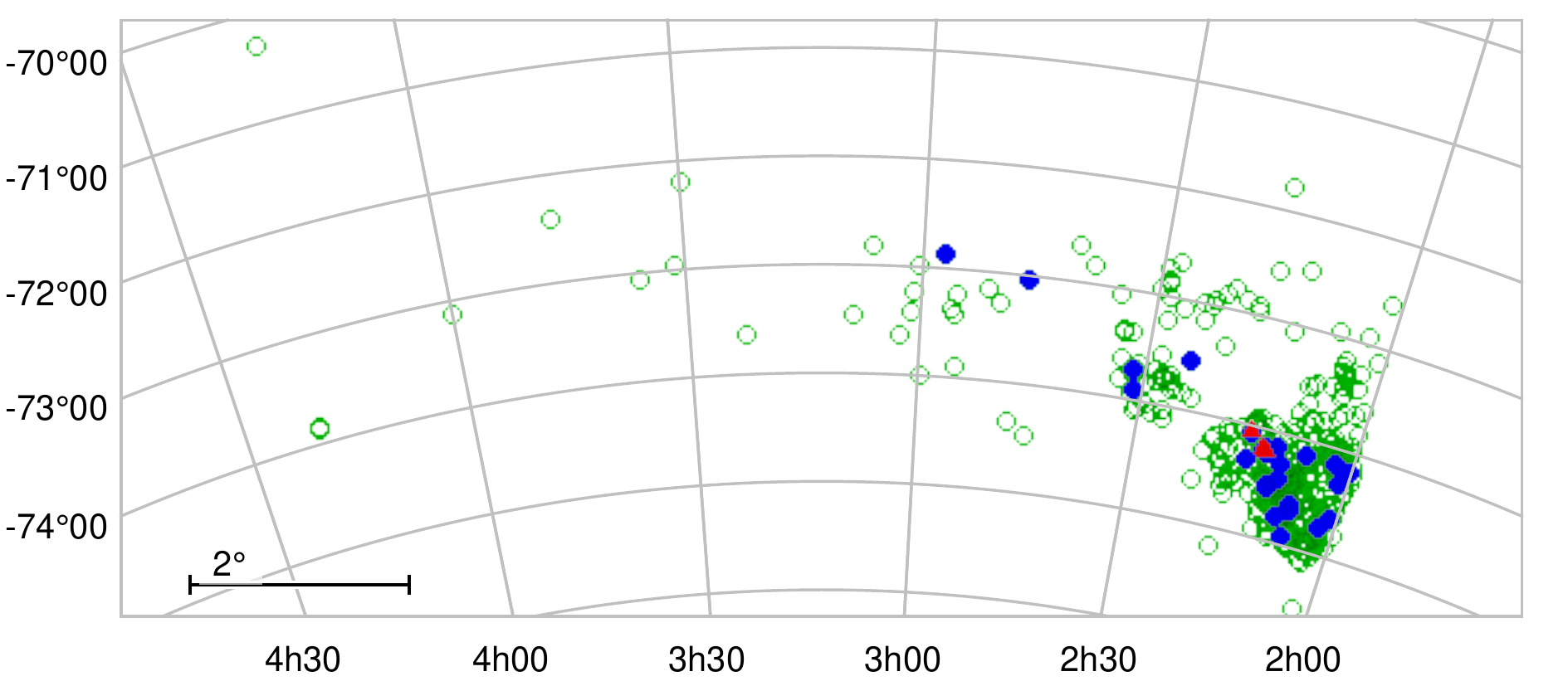}
    \caption{Upper panel: Gaia color–magnitude diagram of Bridge stars which are color coded to their corresponding NUV magnitudes. Our sample O stars are plotted by triangles. Lower panel: Location of OB stars in the Magellanic Bridge (green open circles). Blue filled circles represent the location of stars with $NUV<14.5$, which are candidate O stars. Red triangle shows the location of our analyzed O stars.}
    \label{fig:Gaia}
\end{figure}

To get statistics of hot UV bright stars in the Bridge we use the Gaia DR2 \citep{GaiaDR22018} catalog along with the Galex GR6/7 \citep{Bianchi2017} catalog.
 We select almost all stars in the Gaia catalog which are located between the Magellanic Clouds (as the area shown in the lower panel of Fig.\,\ref{fig:Gaia}).
Subsequently, we apply a magnitude cutoff  $G<17$\,mag. We follow the criterion described in \citet{Zivick2019} to filter out foreground stars by carefully applying a proper motion selection. To find hot stars, we implement a color criterion of $BP-RP<-0.2$\,mag, which results in a total of $\sim 430$ stars. We cross-correlate the final Gaia sample with the Galex GR6/7 catalog. About 410 objects  found to have a Galex $NUV <17$\,mag. Figure\,\ref{fig:Gaia} shows the Gaia color-magnitude diagram (CMD) for the final selected sample of candidate OB stars in the Bridge color-coded with their corresponding NUV magnitude. As expected, stars in the top-left part of the CMD are UV bright. Comparing the position of our analyzed O stars suggests that late O main sequence stars are located near $G\sim15$\,mag, and $BP-RP\sim-0.4$\,mag, with $NUV<14.5$\,mag. Similarly, we found a lower limit for bright O giants from the CMD. Using these criteria, we roughly estimate the total number of O stars in the Bridge to be $\lesssim$25.

With this number of candidate O stars in the Bridge and the $Q_{0}$ values of our analyzed O stars, we estimate the ionization contribution of the whole Bridge massive star population to be $\sim 5 \times 10^{49}$\,ph s$^{-1}$. The O star statistics in the Bridge are comparable to that in the SMC supergiant shell in the Wing \citep{Ramachandran2019}.
Except for the earliest spectral type O3 star, the total flux contribution from the whole OB population in this region is in agreement with our estimated values for the Bridge.

The  observation of the diffuse gas in Bridge by \citet{Lehner2008} suggest an integrated \halpha luminosity $\sim 7 \times 10^{38}$\,erg\,s$^{-1}$. Using the relation $Q(\mathrm{H}\alpha) =7.31 \times 10^{11} L_{\mathrm{H}\alpha}$\,s$^{-1}$ \citep{Kennicutt1995}, we estimate the observed ionizing photon rate to be $<5 \times 10^{50}$\,s$^{-1}$. 
This implies that cumulatively the hot massive stars do not produce sufficient ionizing flux required to explain the H$\alpha$ observation in the Bridge. They only contribute up to $\sim10\%$ of the ionizing flux.

These estimates show that the dominant source of ionization in the Bridge is the hot gas or photon leakage from the Magellanic Clouds. 
The largest and most luminous \hii regions in the Magellanic Clouds are optically thin and located near the edges of these galaxies \citep{Pellegrini2012}. This favors the escape of their ionizing photons to the Bridge. 
Especially the SMC Wing could be an important source of ionization since it is located close to the western part of the Bridge. DEM\,S167 is one of the giant \hii regions in the SMC Wing. \citet{Pellegrini2012} shows that the [\ion{O}{iii}] emission extends to the south, well beyond the ionization transition zone. This provides a blister opening towards the Bridge, resulting in a large photon escape fraction from the region. \citet{Ramachandran2019} estimated the number of escaping ionizing photons from the supergiant shell in the Wing which consists of DEM\,S160 to 167 \hii regions to be  $\sim 10^{50}$\,ph\,s$^{-1}$.  However, the contribution from the SMC Wing alone cannot account for the total ionizing flux observed in the Bridge. 

\citet{Barger2013} noticed a lack of any correlation between the locations of early-type stars in the Bridge and elevated \halpha emission. They suggest that the majority of the ionizing flux comes from the Magellanic Clouds. By studying all known \hii region the Magellanic Clouds, \citet{Pellegrini2012} estimate the total observed \halpha luminosity and lower limits for the escape fraction. Using the relation $L_\mathrm{obs} = L_\mathrm{tot}(1 - f_\mathrm{esc})$, the total ionizing luminosity of the LMC and the SMC are $3\times 10^{40}$\,erg\,s$^{-1}$ and  $4\times 10^{39}$\,erg\,s$^{-1}$, respectively. This translates to an ionizing photon rate of $Q_{0}=2 \times 10^{52}$\,s$^{-1}$ and  $Q_{0}=3\times 10^{51}$\,s$^{-1}$ in the LMC and the SMC. Assuming that the escaping photons from the Magellanic Clouds equally contribute to the ionization of the Bridge, the lower limit on the galactic escape fractions of photons in the LMC and the SMC would be $\sim 1$\% and 8\%, respectively. Even if the ionizing photon contribution from LMC is two or three times higher than from the SMC, 1-2\% of the ionizing photons from the LMC and 4-7\% from the SMC are required to ionize the Bridge.  

\section{Summary}
\label{sec:summary}
\begin{itemize}
    \item We discovered three O-type stars in the Magellanic Bridge. 
This finding demonstrates that despite its low gas density, the Bridge is capable of producing massive O stars.
    \item Spectral analysis reveals that all three stars, MBO1, 2, 3, are likely binaries.
    \item Two stars, MBO2 and MBO3, rotate very slowly, which is uncommon for stars at such low metallicity. We speculate that the tidal interactions of these O stars with their less massive binary companions slowed them down.
    
    \item All three stars have similar age and must have formed during the same star formation event implying on-going star formation in the Bridge. 
    \item The stars are located in the same region of the Bridge, yet they are chemically distinct. MBO1 is as metal-rich as the O stars in the LMC. The mean metallicity of MBO3 is similar to that of previously analyzed B stars in the Bridge. MBO2 is highly CNO deficient, and we claim that this is one of the most metal-poor O stars in the local group.

    \item  We attribute the chemical abundance variations in the Bridge stars to insufficient ISM mixing following multiple gas accretion episodes. 
The \hi gas accreted from the last encounter 200\,Myr ago alone cannot explain the very low metal abundance of MBO2, suggesting that 
the formation of the gaseous component of the Bridge must have started 
many Gyr ago.
    
    \item The ionizing flux from the newly discovered O stars is 
similar to  LMC and SMC stars of the same spectral subtypes.

    \item Using Gaia and Galex data we roughly estimate the total number of O stars in the Bridge and their ionizing flux. 
We show that the total ionizing feedback from the Bridge stars is much lower (< 10\%) than the observed energetics of the diffuse gas. 
We suggest that the ionizing photons escaping from both 
Magellanic Clouds dominate the radiation field in the Bridge.
Based on this assumption we estimate a lower limit for the
escape fractions from the LMC and the SMC to be  1 - 2\% and, 4 - 7\%, respectively.

\end{itemize}

\begin{acknowledgements}
We thank the anonymous referee for very useful
comments and suggestions.
Based on observations collected at the European Organization for
Astronomical Research in the Southern Hemisphere
under ESO programme 078.D-0791 (P.I.: MOMANY, YAZAN).
V.R. is supported by the Deutsches Zentrum für Luft- und Raumfahrt (DLR) under grant 50 OR 1912. This research made use of the VizieR catalog access tool, CDS,  Strasbourg, France. The original description of the VizieR service was published in A\&AS 143, 23. 
\end{acknowledgements}

\bibliographystyle{aa}
\bibliography{main}
 
\appendix


\section{Composite model fits}
\label{sect:appendixa}

\begin{figure*}
    \centering
    \includegraphics[width=\textwidth,trim={0 6cm 0 0}]{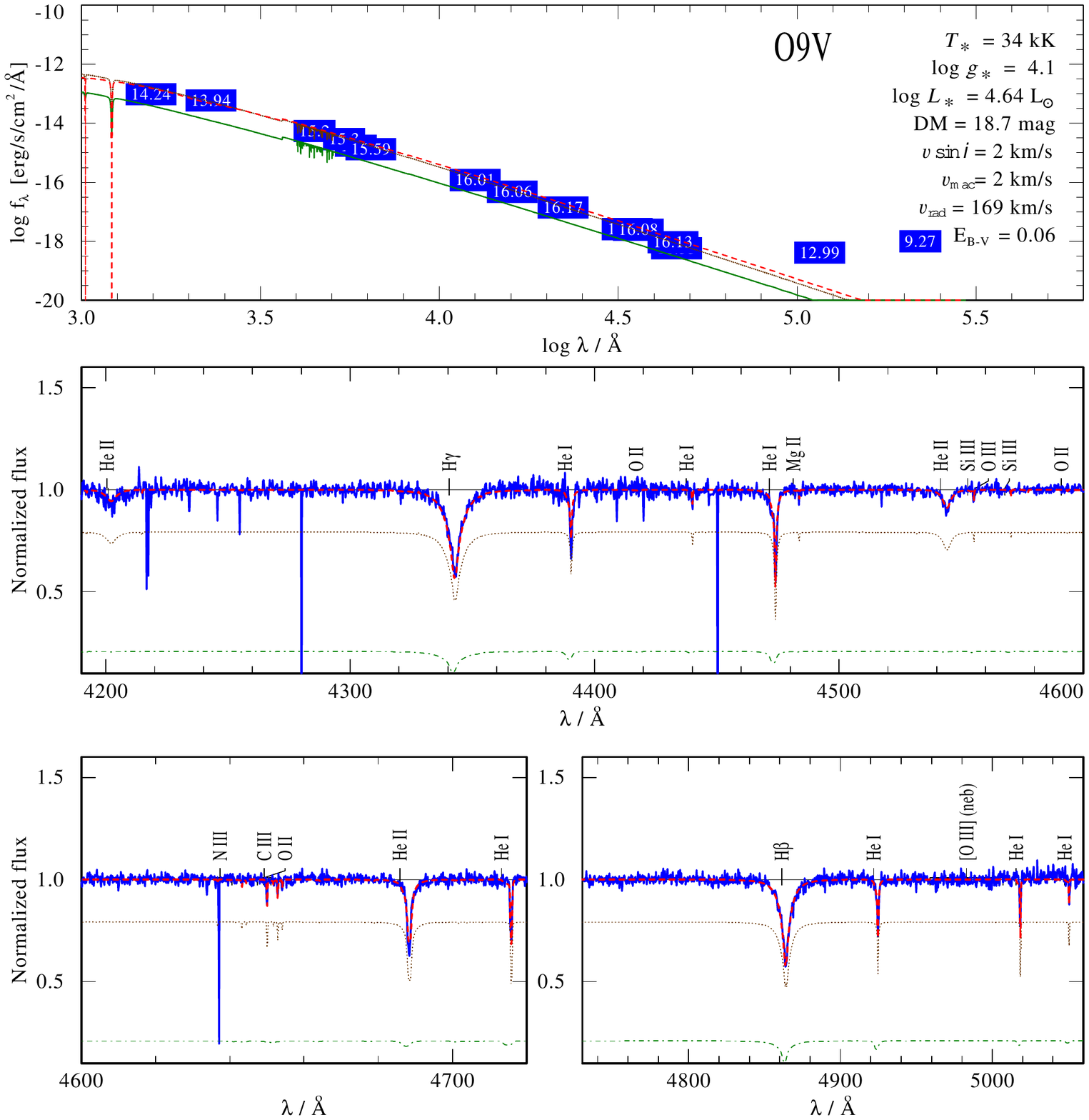}
    \caption{Same as Fig.\,\ref{fig:MBO2}, but a composite model fit (red solid lines) by including an additional B star component (green dashed lines) as secondary. Main parameters of the secondary model are $T _\ast$ = 29\,kK, $\log\,L/L_{\odot}$ =3.92, and	$\log\,g_\ast$ =4.2. We applied faster rotation ($\varv$\,sin\,$i \sim100$\,km\,s$^{-1}$)  and blue ward shift ($\varv_{\rm rad} \sim 110$\,km\,s$^{-1}$) in secondary relative to the primary.
    The primary model (brown dotted lines) is same as that used in the single star fit (see Table\,\ref{table:stellarparameters} and Table\,\ref{table:abundance}). }
    \label{fig:MBO2bin}
\end{figure*}

\begin{figure*}
    \centering
    \includegraphics[width=\textwidth,trim={0 6cm 0 0}]{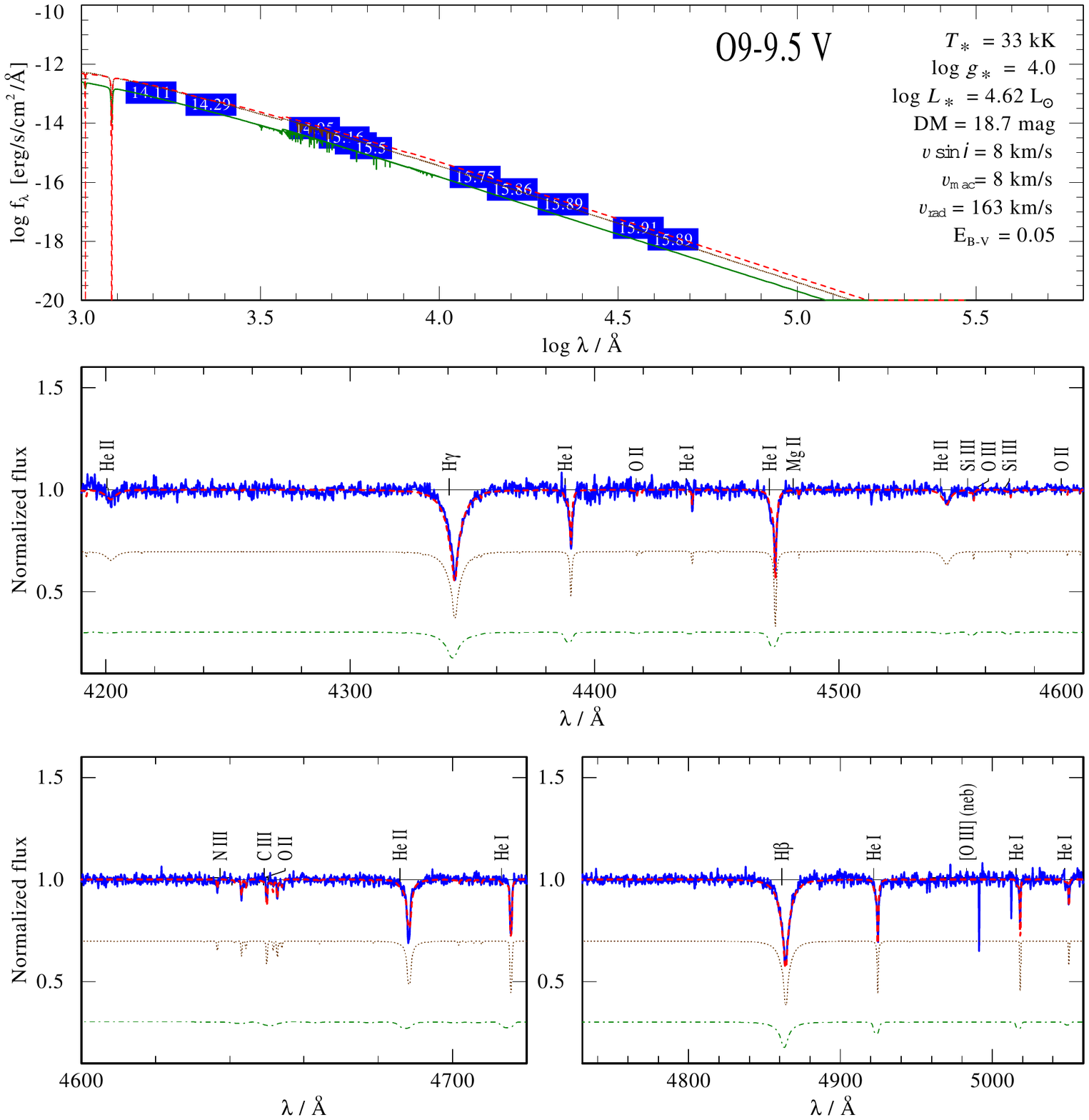}
    \caption{Same as Fig.\,\ref{fig:MBO3}, but a composite model fit (red solid lines) by including an additional B star component (green dashed lines) as secondary.  Main parameters of the secondary model are $T _\ast$ = 31\,kK, $\log\,L/L_{\odot}$ =4.22, and	$\log\,g_\ast$ =4.2. We applied faster rotation ($\varv$\,sin\,$i \sim140$\,km\,s$^{-1}$)  and blue ward shift ($\varv_{\rm rad} \sim 100$\,km\,s$^{-1}$) in secondary relative to the primary.
    The primary model (brown dotted lines) is same as that used in the single star fit (see Table\,\ref{table:stellarparameters} and Table\,\ref{table:abundance}).}
    \label{fig:MBO3bin}
\end{figure*}


\end{document}

%% file: bibdefinitions.tex
\DeclareRobustCommand{\ion}[2]{%
	\relax\ifmmode
	\ifx\testbx\f@series
	{\mathbf{#1\,\mathsc{#2}}}\else
	{\mathrm{#1\,\mathsc{#2}}}\fi
	\else\textup{#1\,{\mdseries\textsc{#2}}}%
	\fi}
\def\degr{\hbox{$^\circ$}}
\def\arcmin{\hbox{$^\prime$}}

\newcommand{\vsini}{\mbox{$\varv \sin i$\ }}

\newcommand{\hi}{\ion{H}{i}\ }
\newcommand{\hii}{\ion{H}{ii}\ }
\newcommand{\hei}{\ion{He}{i}\ }
\newcommand{\heii}{\ion{He}{ii}\ }
\newcommand{\halpha}{H$\alpha$\ }

\def\apjl{ApJ}%
\def\apjs{ApJS}%
\def\ao{Appl.~Opt.}%
\def\aap{A\&A}%